\documentclass[prc,letterpaper,showpacs,12pt,floatfix,preprint]{revtex4}

\usepackage{amsmath}
\usepackage{bm}
\usepackage{graphicx}

\begin{document}
\preprint{\vbox{\hbox{JLAB-THY-09-1107}}}

\title{Ejectile Polarization for $^2  H(e,e'\vec p)n$ at GeV energies}

\author{Sabine Jeschonnek$^{(1)}$ and J. W. Van Orden$^{(2,3)}$}

\affiliation{\small \sl (1) The Ohio State University, Physics
Department, Lima, OH 45804\\
(2) Department of Physics, Old Dominion University, Norfolk, VA
23529\\and\\ (3) Jefferson Lab\footnote{Notice: Authored by Jefferson Science Associates, LLC under U.S. DOE Contract No. DE-AC05-06OR23177. The U.S. Government retains a non-exclusive, paid-up, irrevocable, world-wide license to publish or reproduce this manuscript for U.S. Government purposes.}, 12000 Jefferson Avenue, Newport
News, VA 23606
 }

\date{\today}

\begin{abstract}
We perform a fully relativistic calculation of the $^2  H(e,e'\vec p)n$ reaction in the impulse approximation employing the Gross equation
to describe the deuteron ground state, and we use the SAID parametrization of the full NN scattering amplitude
to describe the final state interactions (FSIs). The formalism for treating the ejectile polarization
with a spin projection on an arbitrary axes is discussed.
We show results for the six relevant asymmetries and discuss the role of spin-dependent FSI contributions.

\end{abstract}

\maketitle

\section{Introduction}

Exclusive electron scattering from the deuteron target is very
interesting by itself, and it also is a very relevant stepping
stone towards understanding exclusive electron scattering from
heavier nuclei. The $D(e,e'p)n$ reaction at GeV energies allows us -
and requires us - to carefully study the reaction mechanism. It is
necessary to consider final state interactions (FSIs) between the
two nucleons in the final state, two-body currents, and isobar
contributions. Of these, the FSIs can be expected to be the most
relevant part of the reaction mechanisms at the GeV energy and
momentum transfers relevant to the study of the transition from
hadronic to quark-gluon degrees of freedom. For some recent reviews
on this exciting topic, see e.g. \cite{wallyreview,ronfranz,sickreview}.

The fact that
the deuteron is the simplest nucleus enables us to study all facets of the reaction mechanism in great detail. Anything
that can be gleaned from the deuteron will be highly useful for heavier nuclei. Exclusive electron scattering from nuclei
is one type of reaction where one may observe color transparency \cite{cteep}, and the deuteron itself provides a laboratory
for the study of neutrons, e.g. the neutron magnetic form factor \cite{hallbgmn}. The short range structures studied
in exclusive electron scattering might even reveal information about the properties of neutron stars \cite{misakneutstar}.

It is important to use all available tools to further our understanding of exclusive scattering from the deuteron.
While unpolarized scattering is interesting, polarization observables hold the promise of revealing more detailed information
about the reaction mechanism. It is therefore important and interesting to test one's model for polarized observables, too.

In \cite{bigdpaper}, we performed a fully relativistic calculation of the $D(e,e'p)n$ reaction, using
a relativistic wave function \cite{wallyfranzwf} and
$NN$ scattering data \cite{said} for our calculation of the full, spin-dependent final state interactions (FSIs).
The main difference to many other high quality calculations  using the generalized
eikonal approximation \cite{misak,ciofi,genteikonal,misakbrandnew} or a diagrammatic
approach \cite{laget} is the inclusion of all the
spin-dependent pieces in the nucleon-nucleon amplitude. Full FSIs have recently been included in \cite{schiavilla}.
 Several experiments with unpolarized deuterons
are currently under analysis or have been published recently, \cite{egiyansrc,halladata,hallbgmn,jerrygexp,blast}.
There are also new proposals for D(e,e'p) experiments at Jefferson Lab \cite{wernernewprop}.

In \cite{bigdpaper}, we focused on observables that are accessible for an unpolarized target and an unpolarized
nucleon detected in the final state. The spin-dependent pieces in our FSI calculation were particularly relevant for
the fifth response function, an observable that can be measured only with polarized electron beams. The spin-dependent
contributions also contributed significantly to the strength in the FSI-dominated regions of the unpolarized cross section.
Naturally, experiments with polarization of the target or ejectile are harder to perform than their unpolarized
counterparts. However, the extra effort allows one to study otherwise inaccessible observables that are rather
sensitive to certain properties of the nuclear ground state and the reaction mechanism. Recently, we investigated
the target polarization in $\vec D(e,e'p)n$ and $\vec D( \vec e,e'p)n$ \cite{targetpol}. In this paper,
we study the asymmetries accessible with a polarized ejectile proton, and a polarized or unpolarized electron beam.
As before, the focus of our
numerical calculations is the kinematic region accessible at GeV energies, i.e. the kinematic range of Jefferson Lab.

Recoil polarization measurement have often been performed for $(e,e' \vec n)$ reactions, to measure the neutron
electric form factor
via polarization transfer. These measurements typically take place at rather low missing momentum. We can perform $D(e,e'\vec n)$
calculations just as easily as $D(e,e'\vec p)$ calculations. For simplicity, we focus on ejectile proton polarization in our
numerical results.

A measurement of two recoil proton polarizations for low missing momentum and various $Q^2$, up to $1.6$ GeV$^2$,
was performed at Jefferson
Lab \cite{hudata}. This was an interesting experiment that used the deuteron as a proton target, and checked if the
deuteron is a good proton target - this is relevant for using the deuteron as a neutron target. At much lower $Q^2$, there
are data from Mainz \cite{mainzrecoil} and Bates \cite{batesdeuteronrecoil} for hydrogen and deuteron targets.
Recoil polarimetry has been used more often for hydrogen targets than deuteron targets, also at Jefferson Lab
\cite{protonelasticrecoil_highqsq,protonelasticrecoil_lowqsq}. Recoil polarimetry continues to be an interesting experimental
technique \cite{fpp_technical}, and it has been used also for heavier nuclei \cite{batescarbonrecoil}, for
photodisintegration \cite{photodisrecoil} and in pion production \cite{kellypionprod}.

This paper is organized as follows: in the next section, we introduce the necessary formalism to define the relevant
observables. In particular, we discuss how to define reduced responses with explicit dependence on the azimuthal angle of
the proton, and we discuss the projection of the proton spin on the most suitable coordinate system.
We define six relevant asymmetries.
Then, we present our numerical results, in a kinematic region relevant to
experiments at Jefferson Lab. We show momentum distributions of all six asymmetries, and we discuss the contributions
of the various spin-dependent parts of the final state interactions.
We conclude with a brief summary.

\section{Formalism}

\subsection{Differential Cross Section}

\begin{figure}
\centerline{\includegraphics[height=3in]{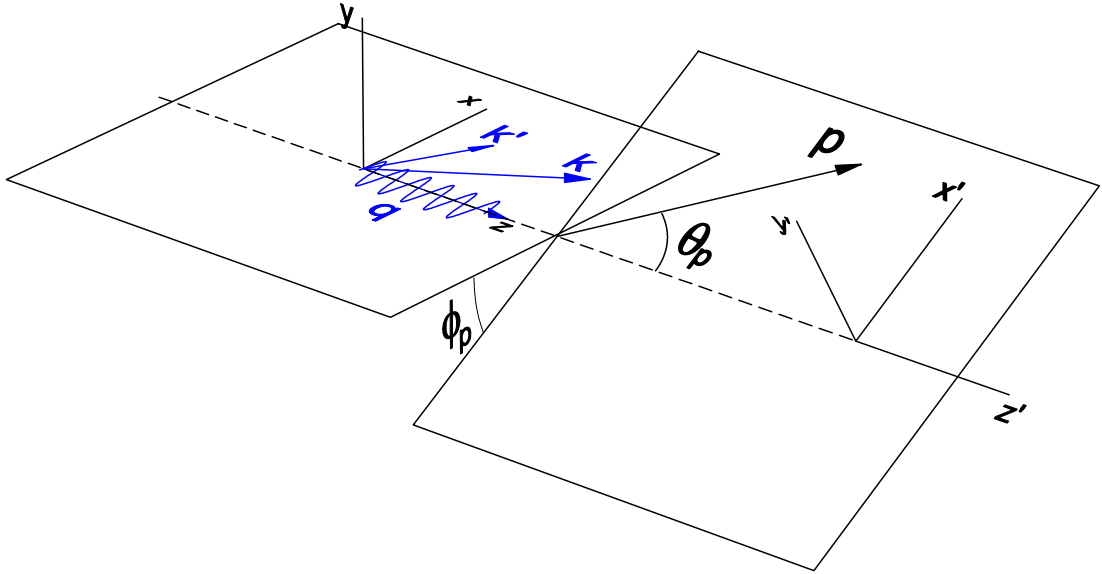}}
\caption{(Color online) Coordinate systems for the $D(e,e'p)$ reaction.  $k$ and $k'$ are the initial and final electron four-momenta,
$q$ is the four-momentum of the virtual photon and $p$ is the four-momentum of the final-state proton.  }\label{coordinates}
\end{figure}

The standard coordinate systems used to describe the $D(e,e'p)$ reaction are shown in Fig.\ref{coordinates}.
The initial and final electron momenta $\bm{k}$ and $\bm{k'}$ define the electron scattering plane
and the $xyz$-coordinate system is defined such that the $z$ axis, the quantization axis, lies along
the momentum of the virtual photon $q$ with the $x$-axis in the electron scattering plane and the $y$-axis
perpendicular to the plane. The momentum $p$ of the outgoing proton is in general not in this plane and is
located relative to the $xyz$ system by the polar angle $\theta_p$ and the azimuthal angle $\phi_p$.
A second coordinate system $x'y'z'$, is chosen such that the $z'$-axis is parallel to the $z$-axis
and the $x'$-axis lies in the hadron plane formed by $\bm{p}$ and $\bm{q}$ and the $y'$-axis is normal to this plane.
The unit vectors in the primed system are defined in terms of the unprimed system as
\begin{eqnarray}
\hat{x}'&=&\cos\phi_p\,\hat{x}+\sin\phi_p\,\hat{y}\nonumber\\
\hat{y}'&=&-\sin\phi_p\,\hat{x}+\cos\phi_p\,\hat{y}\nonumber\\
\hat{z}'&=&\hat{z}\,.
\end{eqnarray}

The general form of the $D(e,e'p)$ cross section can be written in the lab frame as
\cite{raskintwd,dmtrgross}

\begin{eqnarray}
\left ( \frac{ d \sigma^5}{d \epsilon' d \Omega_e d \Omega_p} \right
)_h  & = & \frac{m_p \, m_n \, p_p}{16 \pi^3 \, M_d} \,
\sigma_{Mott} \,
f_{rec}^{-1} \,
 \Big[  v_L R_L +   v_T R_T
 + v_{TT} R_{TT} + v_{LT}R_{LT}
  \nonumber \\
& & +  h \,  v_{LT'} R_{LT'}+h\,v_{T'}R_{T'}
\Big] \, , \label{xsdef}
\end{eqnarray}
where $M_d$, $m_p$ and $m_n$  are the masses of the deuteron, proton and neutron,
 $p_p=p_1$ and $\Omega_p$
are the momentum and solid angle of the ejected proton, $\epsilon'$ is the
energy of the detected electron and $\Omega_e$ is its solid angle, with
$h=\pm 1$  for positive and negative electron helicity. The Mott cross
section is
\begin{equation}
\sigma_{Mott} = \left ( \frac{ \alpha \cos(\theta_e/2)} {2
\varepsilon \sin ^2(\theta_e/2)} \right )^2
\end{equation}
and the recoil factor is given by
\begin{equation}
f_{rec} = \left| 1+ \frac{\omega p_p - E_p q \cos \theta_p} {M_d \, p_p}
\right| \, . \label{defrecoil}
\end{equation}
The leptonic coefficients $v_K$ are
\begin{eqnarray}
v_L&=&\frac{Q^4}{q^4}\\
v_T&=&\frac{Q^2}{2q^2}+\tan^2\frac{\theta_e}{2}\\
v_{TT}&=&-\frac{Q^2}{2q^2}\\
v_{LT}&=&-\frac{Q^2}{\sqrt{2}q^2}\sqrt{\frac{Q^2}{q^2}+\tan^2\frac{\theta_e}{2}}\\
v_{LT'}&=&-\frac{Q^2}{\sqrt{2}q^2}\tan\frac{\theta_e}{2}\\
v_{T'}&=&\tan\frac{\theta_e}{2}\sqrt{\frac{Q^2}{q^2}+\tan^2\frac{\theta_e}{2}}
\end{eqnarray}
Within this general framework, we have two options for evaluating the response functions:
first, we will give expressions for the response functions in terms of matrix elements
that are defined with respect to the electron plane, i.e. the $xyz$ plane. These matrix elements
are implicitly dependent on $\phi_p$, the angle between hadron plane and electron plane, and these are the responses
used e.g. in \cite{raskintwd}.
Second, we give expressions for the responses in the $x'y'z'$ plane. All quantities given relative to the
$x'y'z'$ coordinate system are denoted by a line over the quantity. The current matrix elements, and therefore the
response functions, in the $x'y'z'$ coordinate system do not have any $\phi_p$ dependence. It is much more
practical to evaluate the responses in the $x'y'z'$ coordinate system. The commonly used responses
in the $xyz$ system can then easily be
obtained by accounting for the $\phi_p$ dependence explicitly, see eq.(\ref{connection}) below,
instead of newly evaluating matrix elements for each value of $\phi_p$. Note that both coordinate systems use the same quantization axis: the $z$ axis and the $z'$ axis are parallel.

The hadronic tensor for production of polarized protons is defined as
\begin{equation}
w_{\lambda'_\gamma,\lambda_\gamma}(\hat{{\cal S}})=\frac{2}{3}\sum_{s_1,s'_1,s_2,\lambda_d}
\left<\bm{p}_1s'_1;\bm{p}_2s_2;(-)\right|J_{\lambda'_\gamma}\left|\bm{P}\lambda_d\right>^*
\left<\bm{p}_1s_1;\bm{p}_2s_2;(-)\right|J_{\lambda_\gamma}\left|\bm{P}\lambda_d\right>
{\cal P}_{s'_1s_1}(\hat{{\cal S}})
\end{equation}
where
\begin{equation}
J_{\pm 1}=\mp\frac{1}{\sqrt{2}}(J^1\pm J^2)
\end{equation}
and
\begin{equation}
J_0=J^0
\end{equation}
is the charge operator. The notation $(-)$ in the final state indicates that the state satisfies the boundary conditions appropriate for an ``out'' state. The operator
\begin{equation}
\bm{{\cal P}}(\hat{{\cal S}})=\frac{1}{2}\left(\bm{1}+\bm{\sigma}\cdot \hat{{\cal S}}
\right)
\end{equation}
is a spin projection operator that projects the proton spin onto unit vector $\hat{{\cal S}}$ which corresponds to the direction of the proton spin in its rest frame relative to the $xyz$-system.

The response functions in the $xyz$-frame are given by
\begin{eqnarray}
R_L(\hat{{\cal S}})&=&w_{00}(\hat{{\cal S}})\nonumber\\
R_T(\hat{{\cal S}})&=&w_{11}(\hat{{\cal S}})+w_{-1-1}(\hat{{\cal S}})\nonumber\\
R_{TT}(\hat{{\cal S}})&=&2\mbox{Re}(w_{1-1}(\hat{{\cal S}}))\nonumber\\
R_{LT}(\hat{{\cal S}})&=&-2\mbox{Re}(w_{01}(\hat{{\cal S}})-w_{0-1}(\hat{{\cal S}}))\nonumber\\
R_{LT'}(\hat{{\cal S}})&=&-2\mbox{Re}(w_{01}(\hat{{\cal S}})+w_{0-1}(\hat{{\cal S}}))\nonumber\\
R_{T'}(\hat{{\cal S}})&=&w_{11}(\hat{{\cal S}})-w_{-1-1}(\hat{{\cal S}}) \,.
\label{respdefxyz}
\end{eqnarray}

Now we proceed to write down expressions for the responses in the $x'y'z'$ coordinate system. Calculating
the responses in this system offers a faster alternative to the above calculation, which requires
a new evaluation of the current matrix elements for each $\phi_p$ value.
The response functions defined above are implicitly dependent upon the angle $\phi_p$ between the electron plane and the hadron plane containing the proton and neutron in the final state.  This dependence can be made explicit by noting that
\begin{equation}
\left<\bm{p}_1s_1;\bm{p}_2s_2;(-)\right|J_{\lambda_\gamma}\left|\bm{P}\lambda_d\right>=e^{i(\lambda_d+\lambda_\gamma-s_1-s_2)\phi_p}
\overline{\left<\bm{p}_1s_1;\bm{p}_2s_2;(-)\right|J_{\lambda_\gamma}\left|\bm{P}\lambda_d\right>}
\end{equation}
where the line over the matrix elements is used to indicate that they are quantized relative to the $x'y'z'$ coordinate system. The hadronic tensor can then be written as
\begin{equation}
w_{\lambda'_\gamma,\lambda_\gamma}(\hat{{\cal S}})=e^{-i(\lambda'_\gamma-\lambda_\gamma)\phi_p}\overline{w}_{\lambda'_\gamma,\lambda_\gamma}(\overline{\hat{{\cal S}}})
\label{connection}
\end{equation}
where
\begin{equation}
\overline{w}_{\lambda'_\gamma,\lambda_\gamma}(\overline{\hat{{\cal S}}})=\frac{2}{3}\sum_{s_1,s'_1s_2,\lambda_d}
\overline{\left<\bm{p}_1s'_1;\bm{p}_2s_2;(-)\right|J_{\lambda'_\gamma}\left|\bm{P}\lambda'_d\right>}^*
\overline{\left<\bm{p}_1s_1;\bm{p}_2s_2;(-)\right|J_{\lambda_\gamma}\left|\bm{P}\lambda_d\right>}
{\cal P}_{s'_1s_1}(\overline{\hat{{\cal S}}})
\end{equation}
and
\begin{equation}
{\cal P}_{s'_1s_1}(\hat{\overline{{\cal S}}})=e^{i(s'_1-s_1)\phi_p}{\cal P}_{s'_1s_1}(\hat{{\cal S}})
=\frac{1}{2}\left(\bm{1}+\bm{\sigma}\cdot \hat{\overline{{\cal S}}}
\right)_{s'_1s_1}
\end{equation}
is the spin projection operator defined relative to the $x'y'z'$ coordinate system. Note that this can be obtained by simply decomposing the unit vector $\hat{{\cal S}}$ in terms of the $x'y'z'$ basis.

Using eq. (\ref{connection}) and the definition of the responses in the $xyz$ system, eq. (\ref{respdefxyz}),
the response functions in the $x'y'z'$ system then become
\begin{eqnarray}
R_L(\hat{\overline{{\cal S}}})&=&\overline{R}_L^{(I)}(\hat{\overline{{\cal S}}})\nonumber\\
R_T(\hat{\overline{{\cal S}}})&=&\overline{R}_T^{(I)}(\hat{\overline{{\cal S}}})\nonumber\\
R_{TT}(\hat{\overline{{\cal S}}})&=&\overline{R}_{TT}^{(I)}(\hat{\overline{{\cal S}}})\cos 2\phi_p+\overline{R}_{TT}^{(II)}(\hat{\overline{{\cal S}}})\sin 2\phi_p\nonumber\\
R_{LT}(\hat{\overline{{\cal S}}})&=&\overline{R}_{LT}^{(I)}(\hat{\overline{{\cal S}}})\cos\phi_p+\overline{R}_{LT}^{(II)}(\hat{\overline{{\cal S}}})\sin\phi_p\nonumber\\
R_{LT'}(\hat{\overline{{\cal S}}})&=&\overline{R}_{LT'}^{(I)}(\hat{\overline{{\cal S}}})\sin\phi_p+\overline{R}_{LT'}^{(II)}(\hat{\overline{{\cal S}}})\cos\phi_p\nonumber\\
R_{T'}(\hat{\overline{{\cal S}}})&=&\overline{R}_{T'}^{(II)}(\hat{\overline{{\cal S}}})\label{respdefbar}
\end{eqnarray}
where the reduced response functions for the two classes I and II are defined in terms of the hadronic tensors as
\begin{eqnarray}
\overline{R}^{(I)}_L(\hat{\overline{{\cal S}}})&=& \overline{w}_{00}(\hat{\overline{{\cal S}}}) \nonumber \\
\overline{R}^{(I)}_T(\hat{\overline{{\cal S}}})&=&\overline{w}_{1,1}(\hat{\overline{{\cal S}}})+\overline{w}_{-1,-1}(\hat{\overline{{\cal S}}})  \nonumber \\
\overline{R}^{(I)}_{TT}(\hat{\overline{{\cal S}}})&=&2\mbox{Re}(\overline{w}_{1,-1}(\hat{\overline{{\cal S}}}))  \nonumber \\
\overline{R}^{(II)}_{TT}(\hat{\overline{{\cal S}}})&=& 2\mbox{Im}(\overline{w}_{1,-1}(\hat{\overline{{\cal S}}}))\nonumber \\
\overline{R}^{(I)}_{LT}(\hat{\overline{{\cal S}}})&=&-2\mbox{Re}(\overline{w}_{01}(\hat{\overline{{\cal S}}})-\overline{w}_{0-1}(\hat{\overline{{\cal S}}})) \nonumber   \\
\overline{R}^{(II)}_{LT}(\hat{\overline{{\cal S}}})&=&2\mbox{Im}(\overline{w}_{01}(\hat{\overline{{\cal S}}})+\overline{w}_{0-1}(\hat{\overline{{\cal S}}}))\nonumber\\
 \overline{R}^{(I)}_{LT'}(\hat{\overline{{\cal S}}})&=&2\mbox{Im}(\overline{w}_{01}(\hat{\overline{{\cal S}}})-\overline{w}_{0-1}(\hat{\overline{{\cal S}}})) \nonumber\\
R^{(II)}_{LT'}(\hat{\overline{{\cal S}}})&=& -2\mbox{Re}(\overline{w}_{01}(\hat{\overline{{\cal S}}})+\overline{w}_{0-1}(\hat{\overline{{\cal S}}}))\nonumber\\
\overline{R}^{(II)}_{T'}(\hat{\overline{{\cal S}}})&=& \overline{w}_{1,1}(\hat{\overline{{\cal S}}})-\overline{w}_{-1,-1}(\hat{\overline{{\cal S}}})\, . \label{defresp}
\end{eqnarray}
The reduced response functions still retain some implicit $\phi$ dependence associated with the choice of the arbitrary direction of $\hat{{\cal S}}$ which is fixed relative to the $xyz$ system. This can be made explicit by defining a new set of unit vectors,
usually referred to as normal, longitudinal, and sideways,
\begin{eqnarray}
\hat{n}&=&\hat{y}'\\
\hat{l}&=&\sin\theta_p\,\hat{x}'+\cos\theta_p\,\hat{z}'\\
\hat{s}&=&\cos\theta_p\,\hat{x}'-\sin\theta_p\,\hat{z}'
\end{eqnarray}
fixed relative to the $x'y'z'$ system such that $\hat{l}$ lies along the proton direction with $\hat{s}$ is in the hadron plane and $\hat{n}$ is normal to it. We decompose the spin-dependent part of the projection operator as
\begin{equation}
\bm{\sigma}\cdot\hat{\overline{{\cal S}}}=\bm{\sigma}\cdot\hat{n}\ \hat{n}\cdot\hat{\overline{{\cal S}}}
+\bm{\sigma}\cdot\hat{l}\ \hat{l}\cdot\hat{\overline{{\cal S}}}
+\bm{\sigma}\cdot\hat{s}\ \hat{s}\cdot\hat{\overline{{\cal S}}}
\end{equation}
The response functions can be further expanded as
\begin{equation}
\overline{R}^{(I)}_{K}(\hat{\overline{{\cal S}}})=\overline{R}_K(\bm{1})+\overline{R}_K(\bm{\sigma}\cdot\hat{n})\hat{n}\cdot\hat{\overline{{\cal S}}}
\end{equation}
and
\begin{equation}
\overline{R}^{(II)}_{K}(\hat{\overline{{\cal S}}})=\overline{R}_K(\bm{\sigma}\cdot\hat{l})\hat{l}\cdot\hat{\overline{{\cal S}}}
+\overline{R}_K(\bm{\sigma}\cdot\hat{s})\hat{s}\cdot\hat{\overline{{\cal S}}}
\end{equation}
where the response functions $\overline{R}_K(\cal{O})$, ${\cal O}\in\left\{\bm{1},\bm{\sigma}\cdot\hat{n},\bm{\sigma}\cdot\hat{l},\bm{\sigma}\cdot\hat{s}\right\}$,
can be obtained from (\ref{defresp}) using the response tensors
\begin{equation}
\overline{w}_{\lambda'_\gamma,\lambda_\gamma}({\cal O})=\frac{1}{3}\sum_{s_1,s'_1,s_2,\lambda_d}
\overline{\left<\bm{p}_1s'_1;\bm{p}_2s_2;(-)\right|J_{\lambda'_\gamma}\left|\bm{P}\lambda'_d\right>}^*
\overline{\left<\bm{p}_1s_1;\bm{p}_2s_2;(-)\right|J_{\lambda_\gamma}\left|\bm{P}\lambda_d\right>}
{\cal O}_{s'_1s_1}\,,
\end{equation}
The new response functions are now independent of $\phi_p$ with any residual dependence on $\phi_p$ now contained in the inner products $\hat{n}\cdot\hat{\overline{{\cal S}}}$, $\hat{l}\cdot\hat{\overline{{\cal S}}}$ and $\hat{s}\cdot\hat{\overline{{\cal S}}}$.

It is often convenient to use a simplified notation for these new response functions where
\begin{eqnarray}
\overline{R}_K&=&\overline{R}_K(\bm{1})\nonumber\\
\overline{R}^n_K&=&\overline{R}_K(\bm{\sigma}\cdot\hat{n})\nonumber\\
\overline{R}^l_K&=&\overline{R}_K(\bm{\sigma}\cdot\hat{l})\nonumber\\
\overline{R}^s_K&=&\overline{R}_K(\bm{\sigma}\cdot\hat{s})\,,
\end{eqnarray}
and $\overline{R}_K$ corresponds to the unpolarized response.

Note that when $\theta_p$ is either $0$ or $\pi$, the hadron plane, and therefore the angle $\phi_p$, is no longer defined. As a result the cross section at these angles must be independent of the azimuthal angle $\phi_p$. This imposes constraints on the reduced response functions. The constraints can be obtained by writing the response functions $R_K(\hat{\overline{{\cal S}}})$ of (\ref{respdefbar}) for arbitrary $\hat{{\cal S}}$ in terms of the $\overline{R}^I_K$ with the inner products given explicitly as functions of $\theta_p$ and $\phi_p$. Each of the $R_K(\hat{\overline{{\cal S}}})$ can be expanded to lowest order about $\theta_p=0(\pi)$. The result can then be written as a Fourier series in $\phi_p$.  In order for the cross section to be independent of $\phi_p$ at forward and backward angles all of the coefficients of non-constant terms in the Fourier series must vanish in the limit $\theta_p\rightarrow 0(\pi)$. The resulting equations result in constraints on the $\overline{R}^I_K$. From this analysis, the response functions $\overline{R}_L$, $\overline{R}_T$, $\overline{R}^n_{LT}$, $\overline{R}^n_{LT'}$ and $\overline{R}^l_{T'}$ are unconstrained at these angles while $\overline{R}^s_{LT}=\pm\overline{R}^n_{LT}$ and $\overline{R}^s_{LT'}=\mp\overline{R}^n_{LT'}$ for $\theta_p=0,\pi$.  All other response functions must vanish at forward and backward angles.

\subsection{Asymmetries}

The definition of asymmetries for polarized protons must be done carefully.  Experiments to date have been done for the case where $\phi_p=0$.  In this case the asymmetries have been determined relative to the unit vectors $\hat{n}$, $\hat{l}$ and $\hat{s}$.
However, the use of this approach for out-of-plane measurements results in asymmetries that depend on $\phi_p$ when $\theta_p=0,\pi$.
Therefore, our goal is to define a coordinate system that will be defined unambiguously even if $\theta_p=0,\pi$.
It turns out that a reference frame suggested by the experimental set-up fulfills this requirement.

Another approach can be defined by noting that if a magnetic spectrometer is used to detect the proton, out-of-plane angles are most conveniently reached by tilting the spectrometer relative to the lab floor with the horizontal direction in the spectrometer remaining fixed. As a result, a new set of coordinates can be chosen such that the longitudinal axis lies along the direction of the proton and the sideways direction remains parallel to the electron scattering plane. The unit vectors defining this system are then given by
\begin{eqnarray}
\hat{l}'&=&\hat{l}\\
\hat{s}'&=&\frac{\hat{y}\times\hat{l}}{|\hat{y}\times\hat{l}|}\\
\hat{n}'&=&\hat{l}'\times\hat{s}'\,.
\end{eqnarray}

Choosing $\hat{\overline{{\cal S}}}=\hat{n}'$, the unpolarized part of the cross section, which is independent of $\hat{n}'$ can be written as
\begin{equation}
\sigma(0)+h \sigma_h(0)
\end{equation}
and the part of the cross section proportional to $\hat{n}'$ can be written as
\begin{equation}
\sigma(n')+h \sigma_h(n')
\end{equation}
and choosing $\hat{\overline{{\cal S}}}=\hat{l}'$ or $\hat{\overline{{\cal S}}}=\hat{s}'$ can be used to obtain the contributions
\begin{equation}
\sigma(l')+h \sigma_h(l')
\end{equation}
and
\begin{equation}
\sigma(s')+h \sigma_h(s')
\end{equation}

The single and double asymmetries are now defined as
\begin{equation}
\label{asymdef1}
A^\xi_p=\frac{\sigma(\xi)}{\sigma(0)}
\end{equation}
and
\begin{equation}
\label{asymdef2}
A^\xi_{ep}=\frac{\sigma_h(\xi)}{\sigma(0)}
\end{equation}
where $\xi=n',l',s'$. These asymmetries can be shown to be independent of $\phi_p$ for $\theta_p=0,\pi$.

\subsection{Current Matrix Elements}

A detailed description of the impulse approximation current matrix elements used here is presented in \cite{bigdpaper}. These matrix elements are constructed based on the covariant spectator approximation \cite{grosseqn}. A relativistic wave function \cite{wallyfranzwf} and
$NN$ scattering data \cite{said} are used for our calculation of the full, spin-dependent final state interactions.
The main difference to many other high quality calculations  using the generalized
eikonal approximation \cite{misak,misakbrandnew,ciofi,genteikonal} or a diagrammatic
approach \cite{laget} is the inclusion of all the
spin-dependent pieces in the nucleon-nucleon amplitude. Full FSIs have recently been included in \cite{schiavilla}.

To construct the scattering amplitudes needed for the calculation of the FSIs we start with $np$ helicity matrices extracted from SAID \cite{said}. The on-shell scattering amplitudes can be given in terms of five Fermi
invariants as
\begin{eqnarray}
M_{ab;cd}&=&\mathcal{F}_S(s,t)\delta_{ac}\delta_{bd}+\mathcal{F}_V(s,t)\gamma_{ac}
\cdot\gamma_{bd}+\mathcal{F}_T(s,t)\sigma^{\mu\nu}_{ac}(\sigma_{\mu\nu}) _{bd}^{} \nonumber\\
&&+\mathcal{F}_{P}(s,t)\gamma^5_{ac}\gamma^5_{bd}+
\mathcal{F}_A(s,t)(\gamma^5\gamma)_{ac}\cdot(\gamma^5\gamma)_{bd}\label{Fermi}
\label{eqdefnn}
\end{eqnarray}
where $s$ and $t$ are the usual Mandelstam variables. The Fermi invariants are then determined using the helicity amplitudes.
A table of the invariant functions is constructed in terms of $s$ and the center of
momentum angle $\theta$.  The table is then interpolated to obtain the
invariant functions at the values required by the integration.

In order to estimate
the possible effects of this contribution to the current matrix
elements, we use a simple prescription for the off-shell behavior of
the amplitude. Although additional invariants are possible when the
nucleon is allowed to go off shell, we keep only the forms in
(\ref{Fermi}).  The center-of-momentum angle is calculated using

\begin{equation}
\displaystyle{\cos\theta=\frac{t-u}{\sqrt{s-4m^2}\sqrt{\frac{(4m^2-t-u)^2}{s}-4m^2}}}\label{thetacm}
\end{equation}
The invariants are then replaced by
\begin{equation}
\mathcal{F}_i(s,t)\rightarrow\mathcal{F}_i(s,t,u)F_N(s+t+u-3m^2)
\end{equation}
where
\begin{equation}
F_N(p^2)=\frac{(\Lambda_N^2-m^2)^2}{(p^2-m^2)^2+(\Lambda_N^2-m^2)^2}\label{Nff}
\end{equation}
and the $\mathcal{F}_i(s,t,u)$ are obtained from interpolation of
the on-shell invariant functions with the center-of-momentum angle
obtained from (\ref{thetacm}). The form factor (\ref{Nff}) is used as a cutoff to limit contributions where the nucleon is highly off shell. We use a value of $\Lambda_N = 1$ GeV in this paper. The numerical effects of variations in the cut-off parameter have been studied in \cite{bigdpaper}.

\section{Results}

\subsection{Momentum Distributions}

\begin{figure}[ht]
\includegraphics[width=14pc,angle=270]{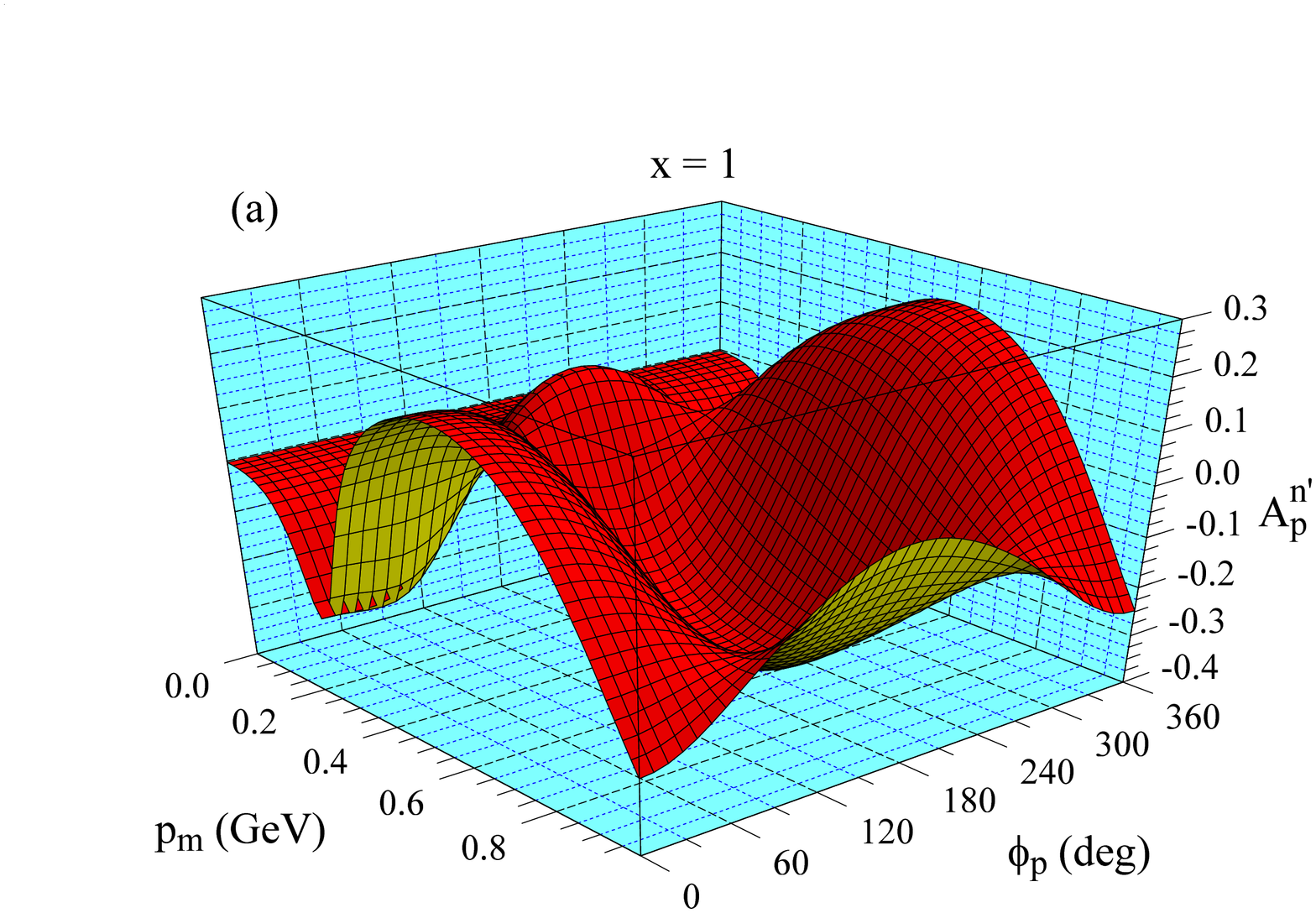}
\includegraphics[width=14pc,angle=270]{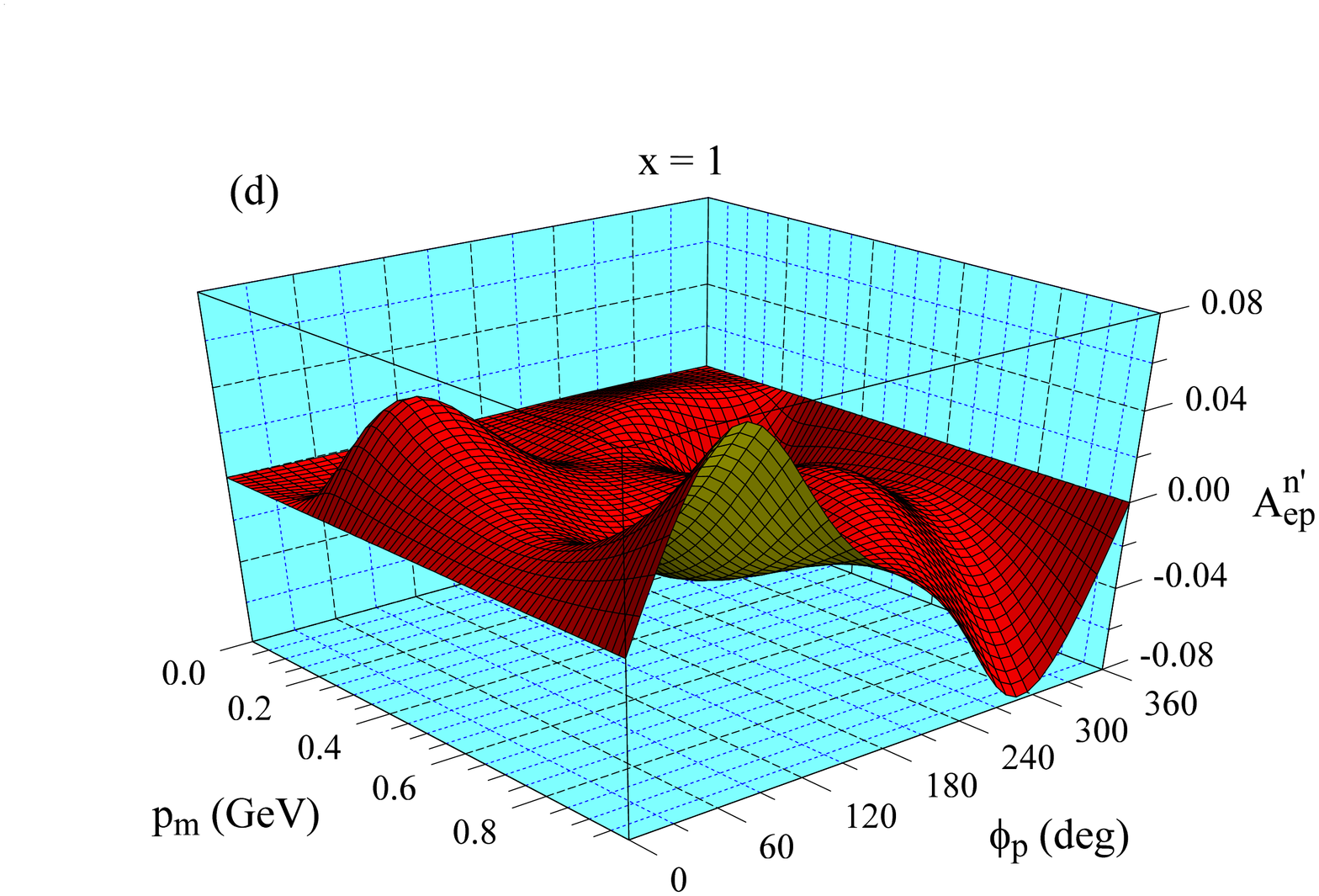}
\includegraphics[width=14pc,angle=270]{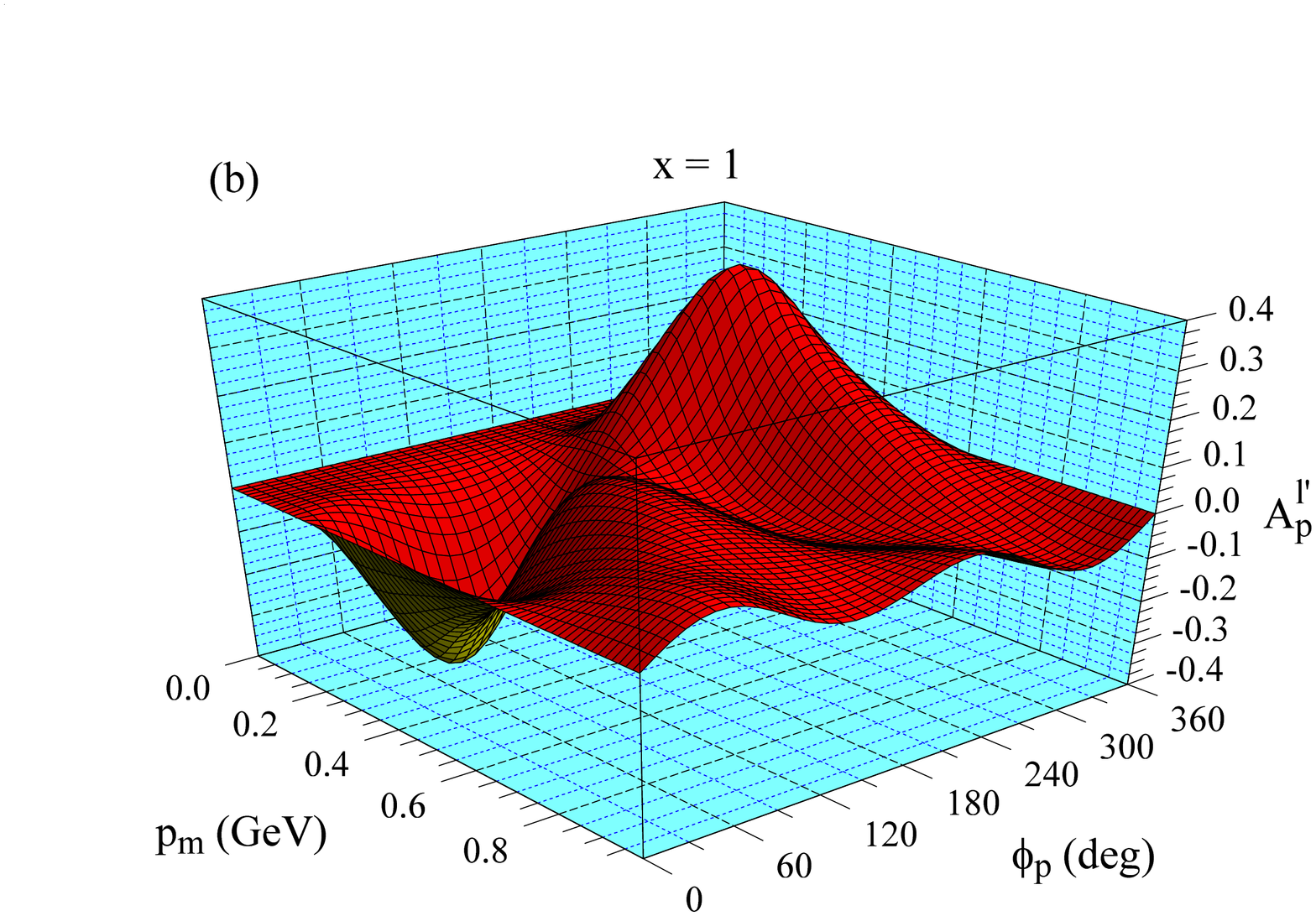}
\includegraphics[width=14pc,angle=270]{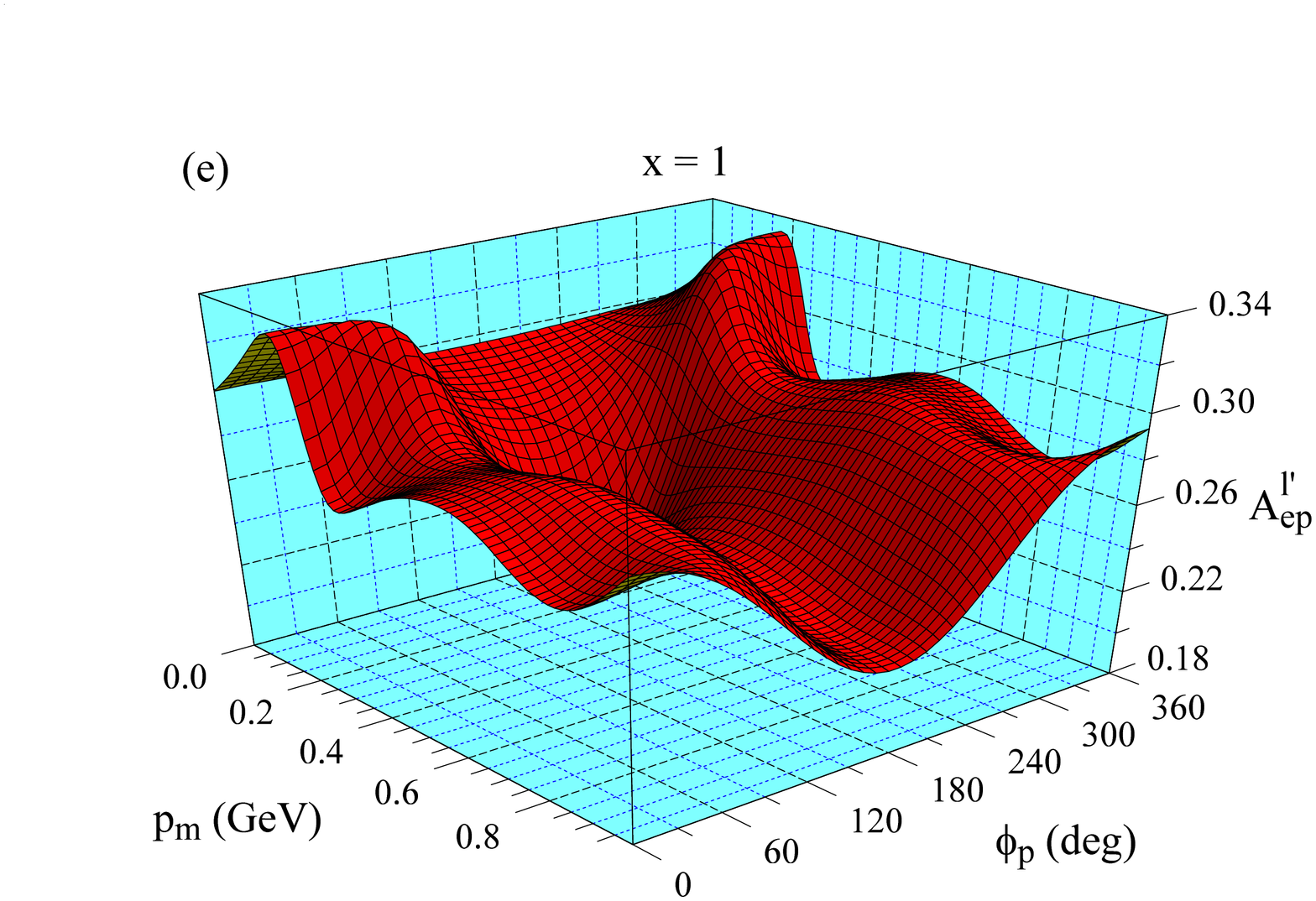}
\includegraphics[width=14pc,angle=270]{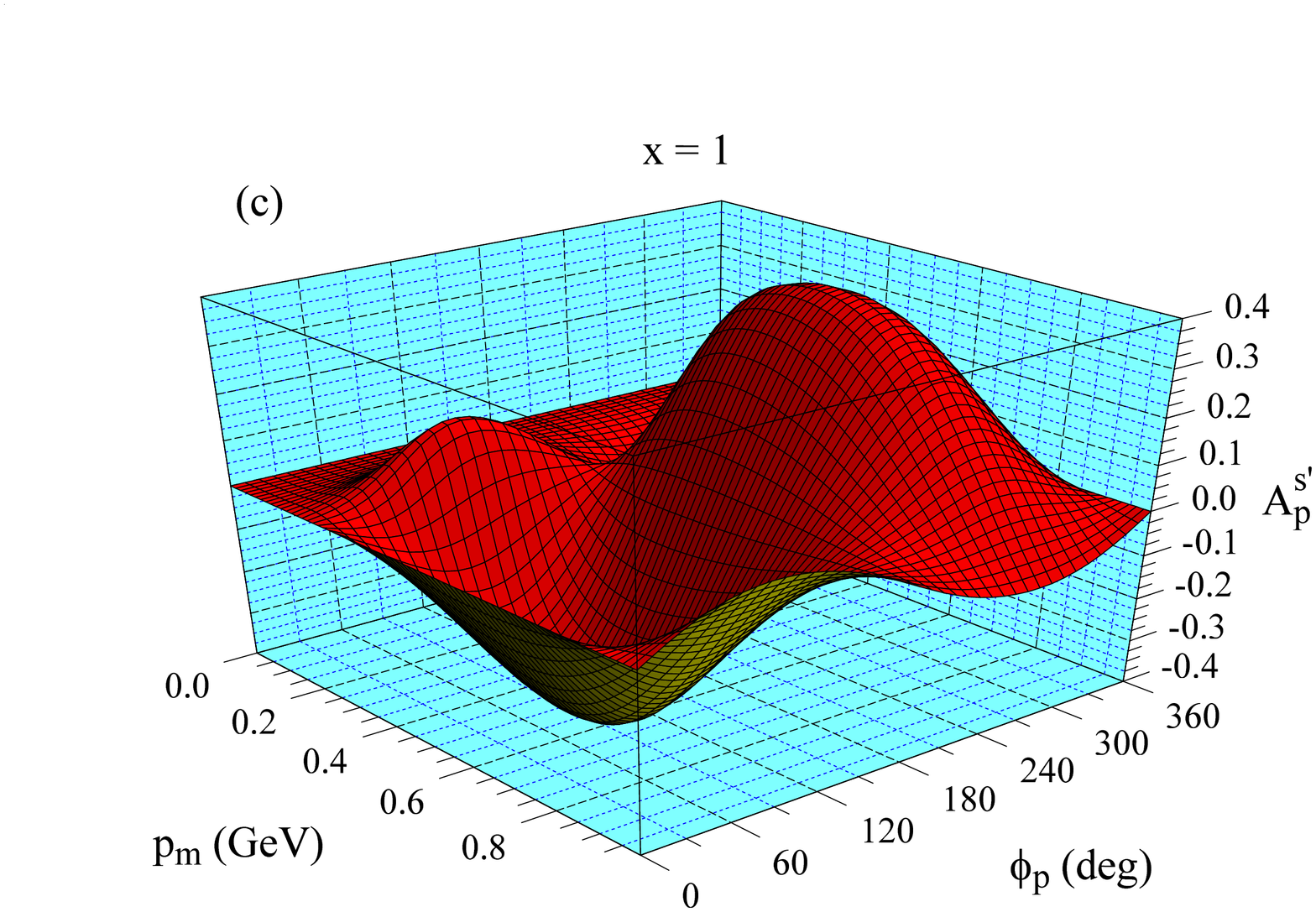}
\includegraphics[width=14pc,angle=270]{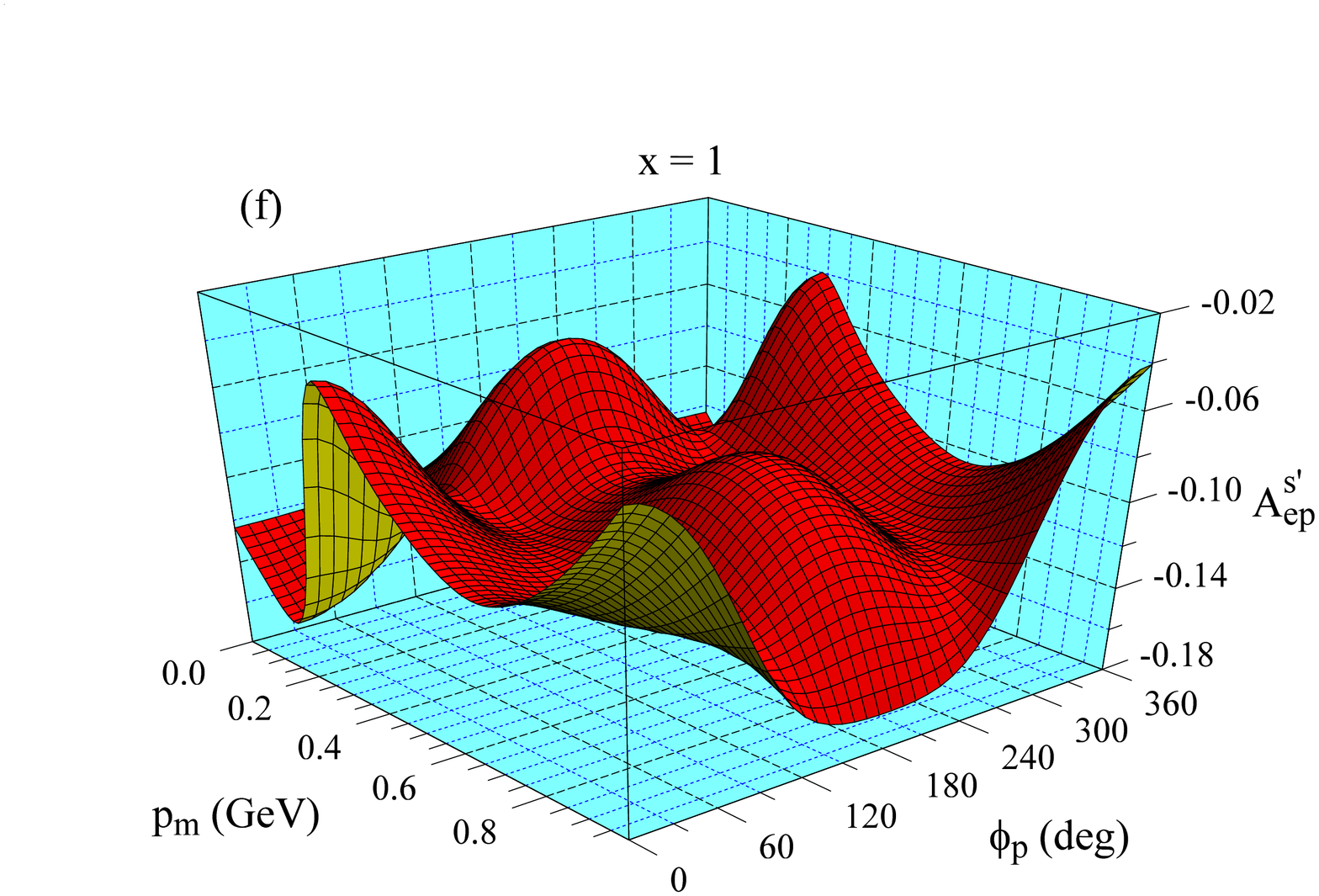}
\caption{(Color online) The six panels show the six asymmetries plotted versus the missing momentum $p_m$ and
versus the azimuthal angle of the proton, $\phi_p$, for a beam energy of $5.5$ GeV, a transferred four-momentum
of $Q^2 = 2$ GeV$^2$, and $x = 1$. We show $A^{n'}_p$ (panel a)), $A^{l'}_p$ (panel b)), $A^{s'}_p$ (panel c)),
$A^{n'}_{ep}$ (panel d)), $A^{l'}_{ep}$ (panel e)), and $A^{s'}_{ep}$ (panel f)). The curves shown have been calculated
including on-shell FSIs.}
\label{fig_md_3d_x1}
\end{figure}

In order to give a general overview of the properties of all six asymmetries, we show them
in Fig. \ref{fig_md_3d_x1} as three-dimensional plots versus the missing momentum and the
azimuthal angle of the proton, $\phi_p$. From these plots, it becomes obvious that any statements
about the relative size of the asymmetries are highly dependent on the independent variables,
and none of the asymmetries can be singled out as ``the largest'' or ``the smallest'' in general.
If one restricts one's interest to in-plane measurements, i.e. to $\phi_p \approx 0^o$, one will observe
that $A^{l'}_{ep}$ is larger than the other observables, and $A^{n'}_p$ and $A^{l'}_p$ are
medium-sized, but this is a $\phi_p$ dependent statement.

Two of the asymmetries, $A^{s'}_p$ and
$A^{n'}_{ep}$, appear to be antisymmetric around $\phi_p = 180^o$, while the other four asymmetries are symmetric
around $\phi_p = 180^o$.

One of the most interesting questions is how large the influence of final state interactions is, both of the on-shell and off-shell
contributions. Three of the asymmetries are zero in PWIA, and so the FSI influence in these cases - for $A^{n'}_p$, $A^{l'}_p$,
and $A^{s'}_p$ - is obviously large. All three asymmetries take medium-size or large values somewhere in the kinematics plane shown
in Fig. \ref{fig_md_3d_x1}.

\begin{figure}[ht]
\includegraphics[width=14pc,angle=270]{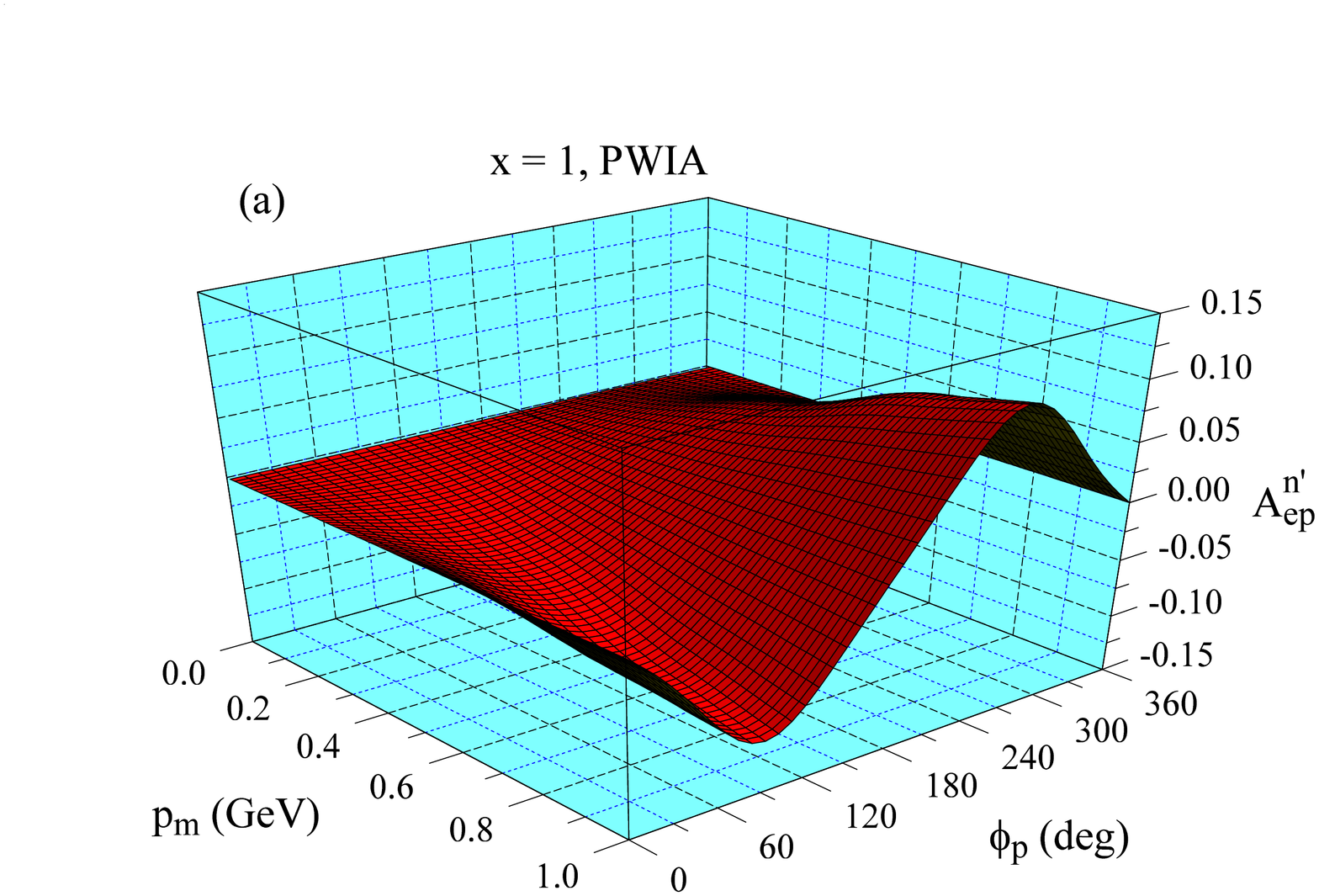}
\includegraphics[width=14pc,angle=270]{anep_3d_x1.ps}
\includegraphics[width=14pc,angle=270]{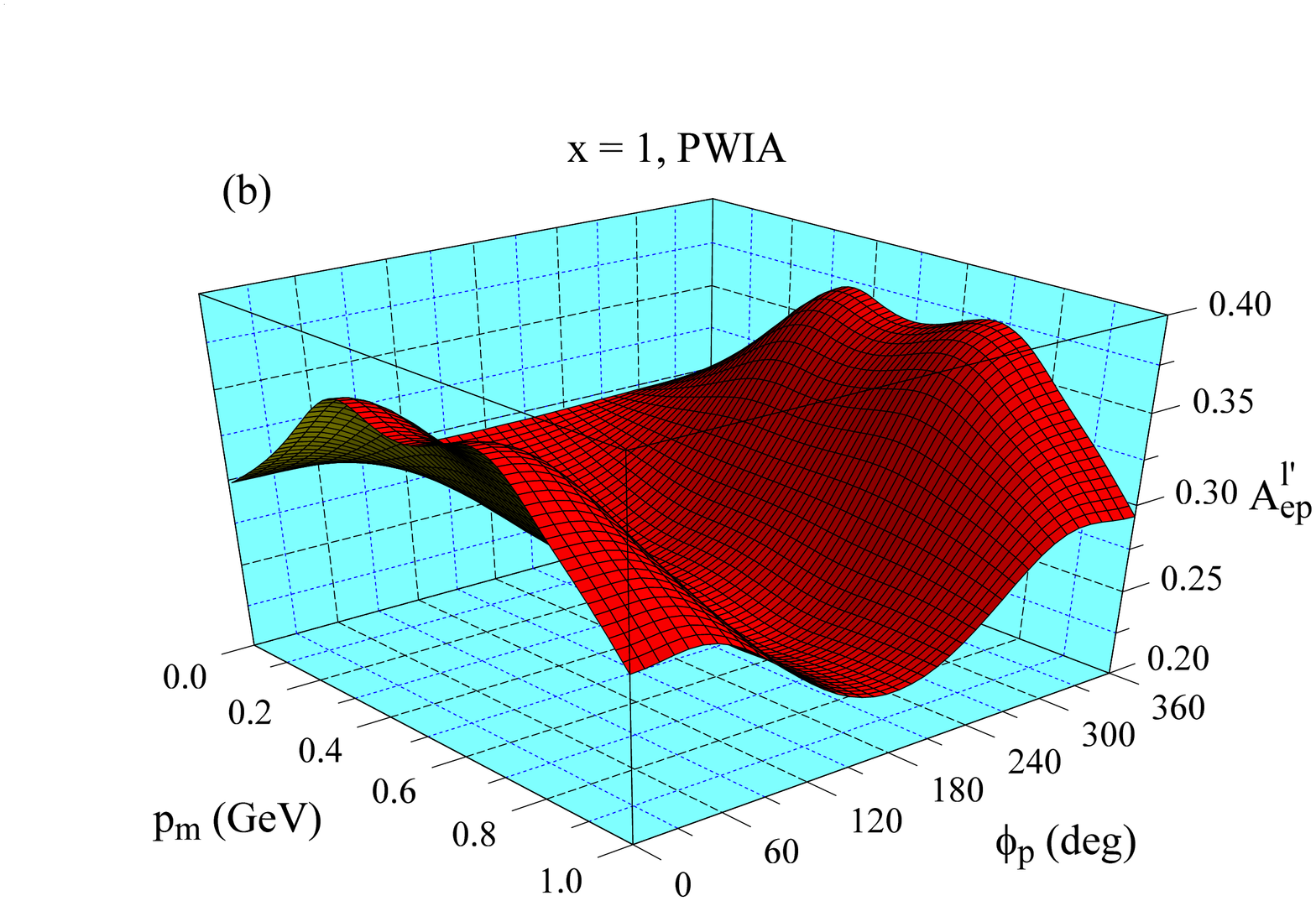}
\includegraphics[width=14pc,angle=270]{alep_3d_x1.ps}
\includegraphics[width=14pc,angle=270]{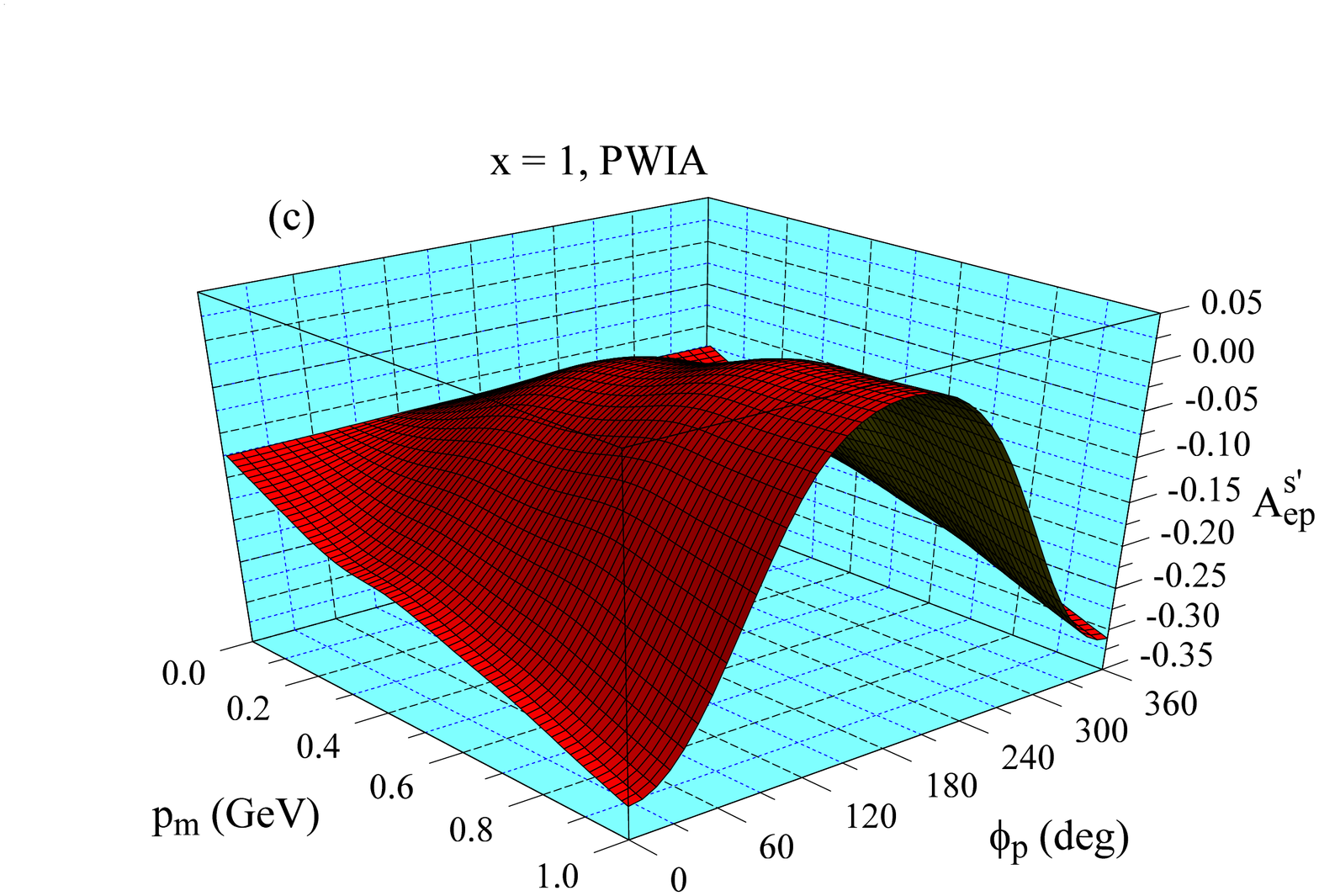}
\includegraphics[width=14pc,angle=270]{asep_3d_x1.ps}
\caption{(Color online) The six panels show the three asymmetries that are non-zero in PWIA plotted versus the missing momentum $p_m$ and
versus the azimuthal angle of the proton, $\phi_p$, for a beam energy of $5.5$ GeV, a transferred four-momentum
of $Q^2 = 2$ GeV$^2$, and $x = 1$. We show the PWIA results in the left column (panels a), b), and c)), and the on-shell FSI
results in the right column (panels d), e), and f)). The top row shows $A^{n'}_{ep}$, the middle row shows $A^{l'}_{ep}$,
and the bottom row shows $A^{s'}_{ep}$. The kinematics are the same as in the previous figure.}
\label{fig_md_3d_x1_pwiafsi}
\end{figure}

In Fig. \ref{fig_md_3d_x1_pwiafsi}, we show three-dimensional plots of the three asymmetries that are non-zero
in PWIA, i.e. the asymmetries that need a polarized electron beam. The left column shows the PWIA results, the right column shows the corresponding results obtained with on-shell FSIs included. Again, it is obvious that the influence of the FSIs is large. FSIs change the
shape and the magnitude of the asymmetries. While any asymmetry can be either drastically increased or decreased at any point of the
covered kinematics, one can see that the overall effect of FSIs is to reduce the asymmetries somewhat.

\begin{figure}[ht]
\includegraphics[width=14pc,angle=270]{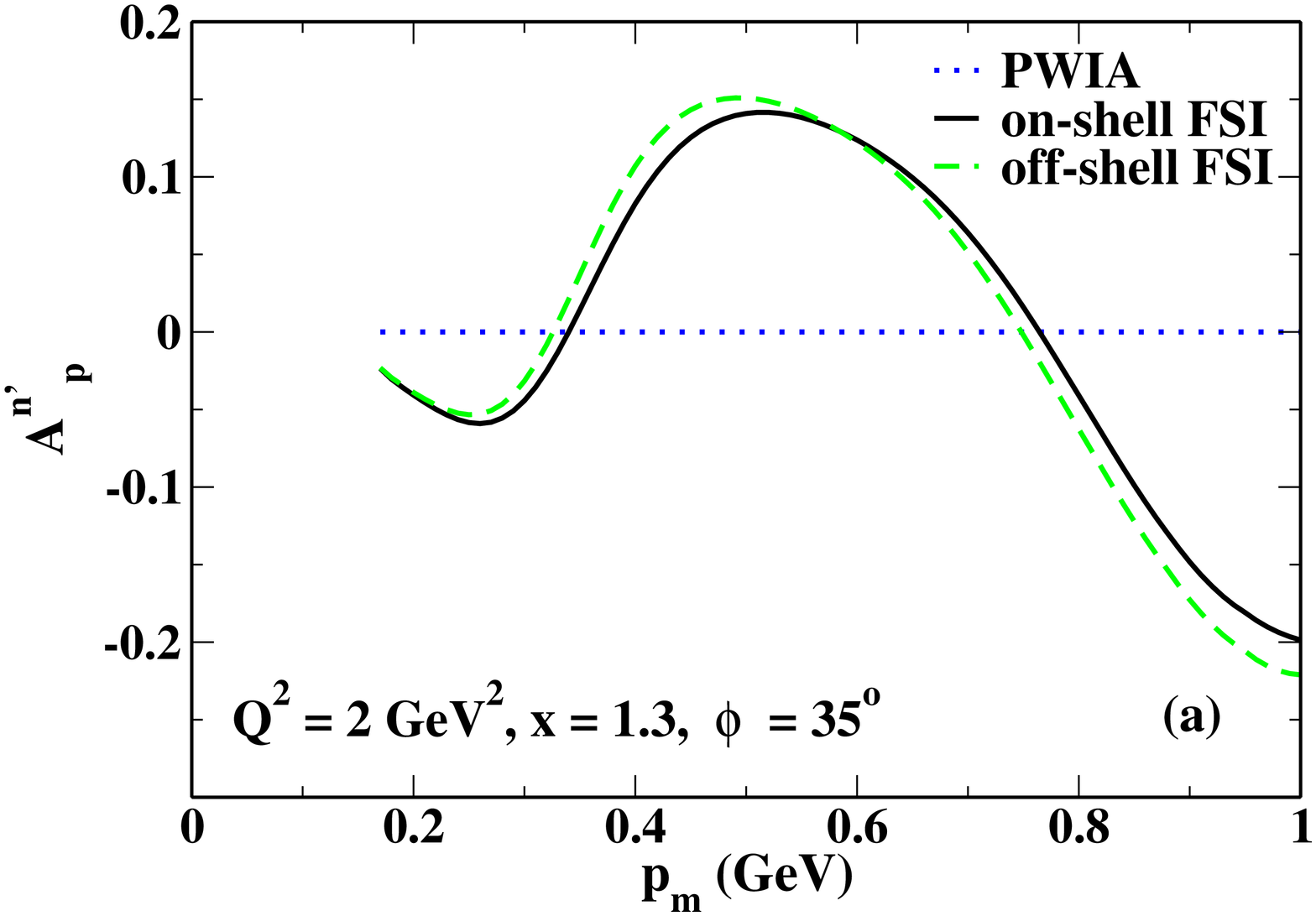}
\includegraphics[width=14pc,angle=270]{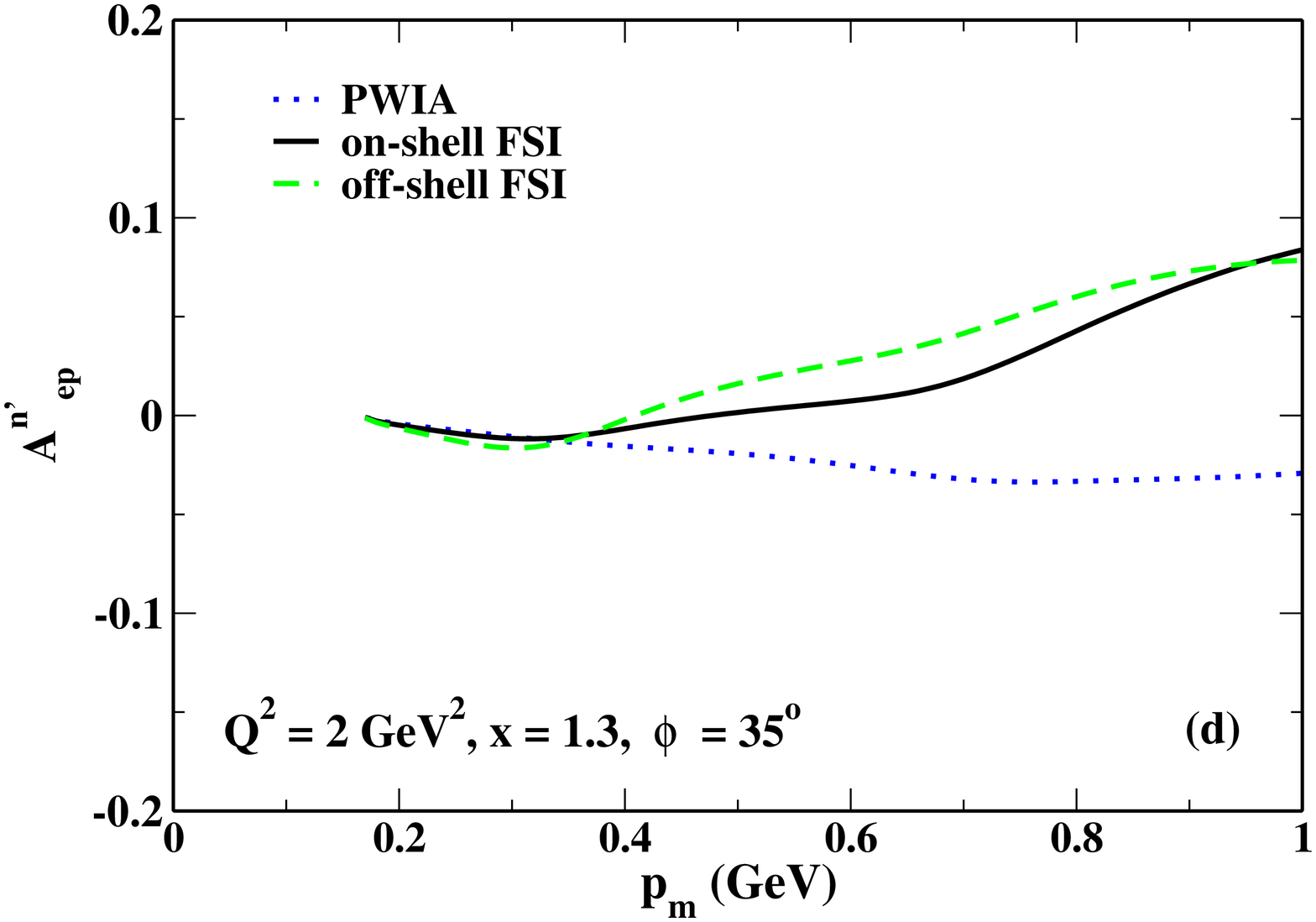}
\includegraphics[width=14pc,angle=270]{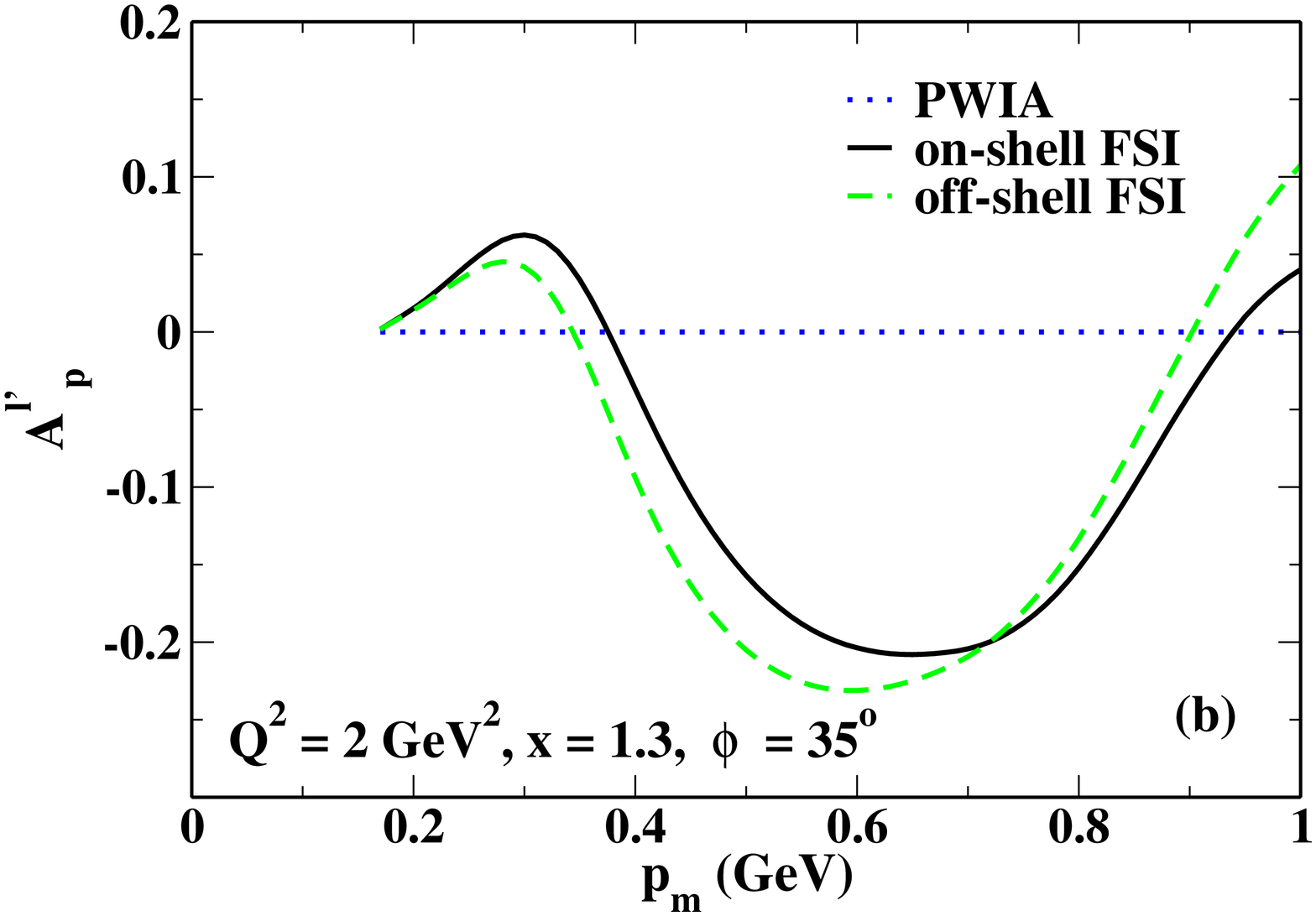}
\includegraphics[width=14pc,angle=270]{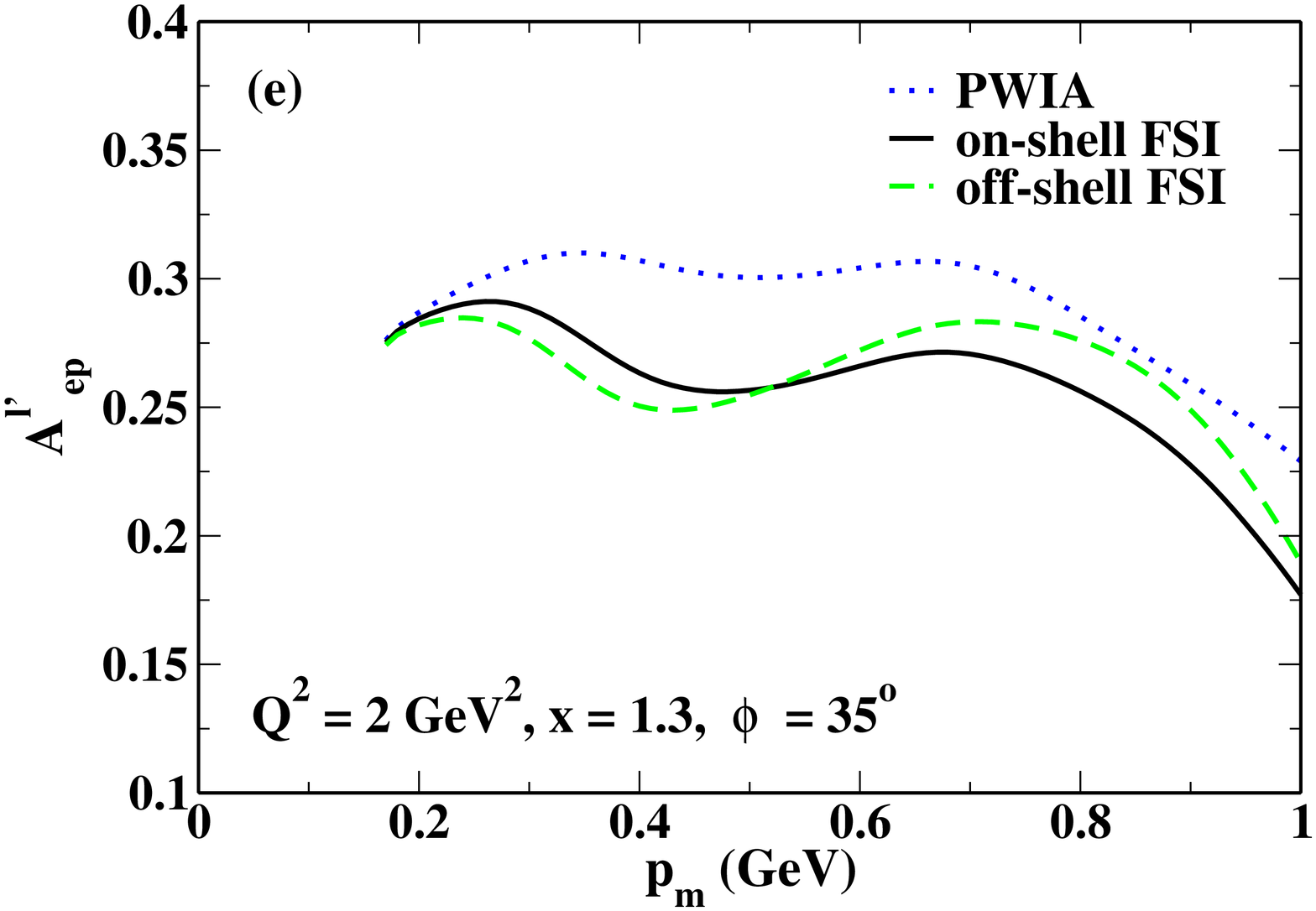}
\includegraphics[width=14pc,angle=270]{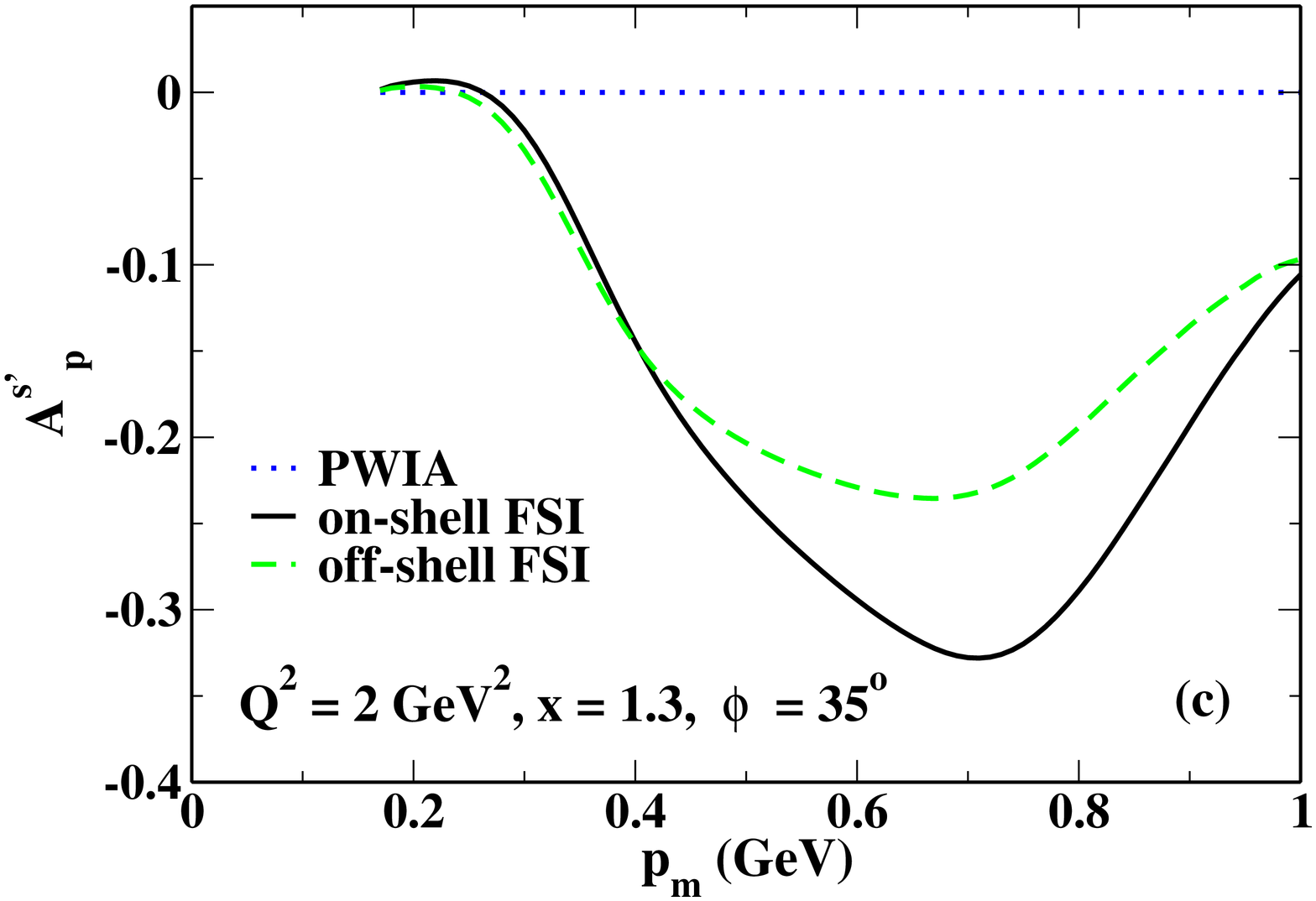}
\includegraphics[width=14pc,angle=270]{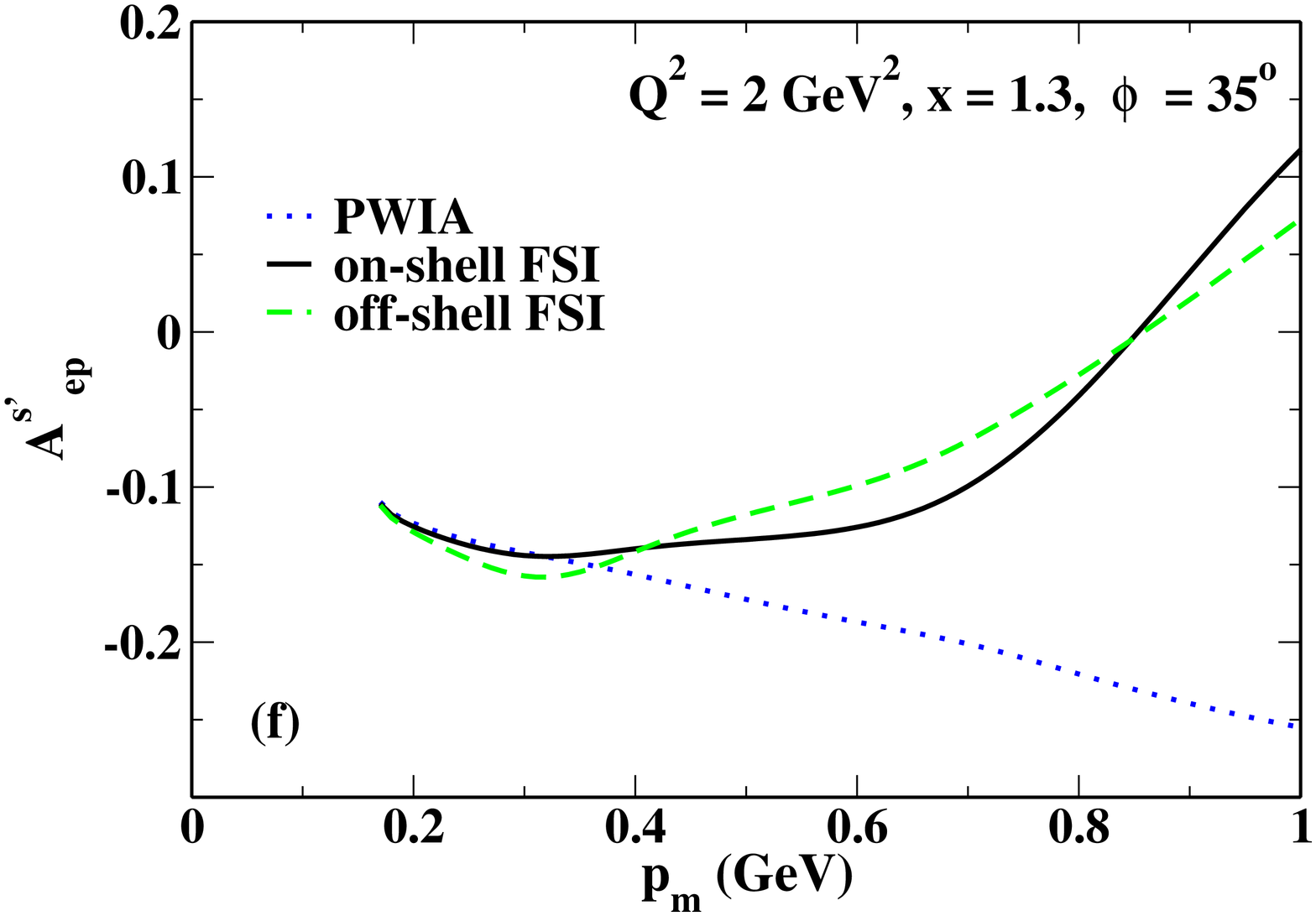}
\caption{(Color online)  The six panels show the six asymmetries plotted versus the missing momentum $p_m$
for a beam energy of $5.5$ GeV, a transferred four-momentum
of $Q^2 = 2$ GeV$^2$, $\phi_p = 35^o$, and $x = 1.3$. We show $A^{n'}_p$ (panel a)), $A^{l'}_p$ (panel b)), $A^{s'}_p$ (panel c)),
$A^{n'}_{ep}$ (panel d)), $A^{l'}_{ep}$ (panel e)), and $A^{s'}_{ep}$ (panel f)). The curves shown have been calculated
in PWIA (dotted), with on-shell FSIs (solid), and including off-shell and on-shell FSIs (dashed). }
\label{fig2dx13}
\end{figure}

We will now turn to the discussion of the off-shell contribution to the FSIs. For $x = 1$, the quasi-elastic region, they turn out to
be fairly small. This is what we expect, and what we have observed earlier, for the unpolarized case \cite{bigdpaper}
and for a polarized deuteron target \cite{targetpol}. Therefore, we do not display results for $x = 1$, but we move away from the
quasi-elastic region, to $x = 1.3$, where the off-shell contributions to the FSIs should be a bit larger.
In Fig. \ref{fig2dx13}, we show two-dimensional plots of
the six asymmetries. The PWIA contribution is shown as the dotted line, the on-shell FSIs are shown by the solid line, and the
calculation including the off-shell FSIs is shown by the dashed line. It is again easy to see that FSIs are very important. It also turns
out that the off-shell FSIs lead only to modest corrections, and they never lead to qualitative changes in the shape of an asymmetry.
The largest effects can be seen in $A^{s'}_{p}$, where the value of the asymmetry is reduced significantly for large missing
momenta around $p_m = 0.7$ GeV. For $A^{n'}_{ep}$, there is a noticeable increase due to the off-shell FSIs.

\subsection{Contributions from Individual Parts of the $pn$ scattering amplitude to the FSIs}

In our calculation of the final state interactions, we use the full nucleon-nucleon scattering amplitude.
There are several ways to decompose and parametrize the
$NN$ scattering amplitude. It can be parametrized with five terms: a central, spin-independent term, a spin-orbit term, and three
double-spin flip contributions. It can also be given in terms of invariants, using a scalar, vector, tensor, pseudoscalar, and axial
term. Some of these parametrizations may be more or less useful and enlightening in trying to understand what is happening.
As we are interested in the ejectile polarization, investigating the effects of spin-dependent terms in the FSIs
is a logical and interesting step. We separate the NN amplitudes into a central term, a single spin-flip (i.e. spin-orbit) term,
and three double spin-flip terms, for details on these Saclay amplitude conventions, see \cite{bigdpaper}.

\begin{figure}[ht]
\includegraphics[width=14pc,angle=270]{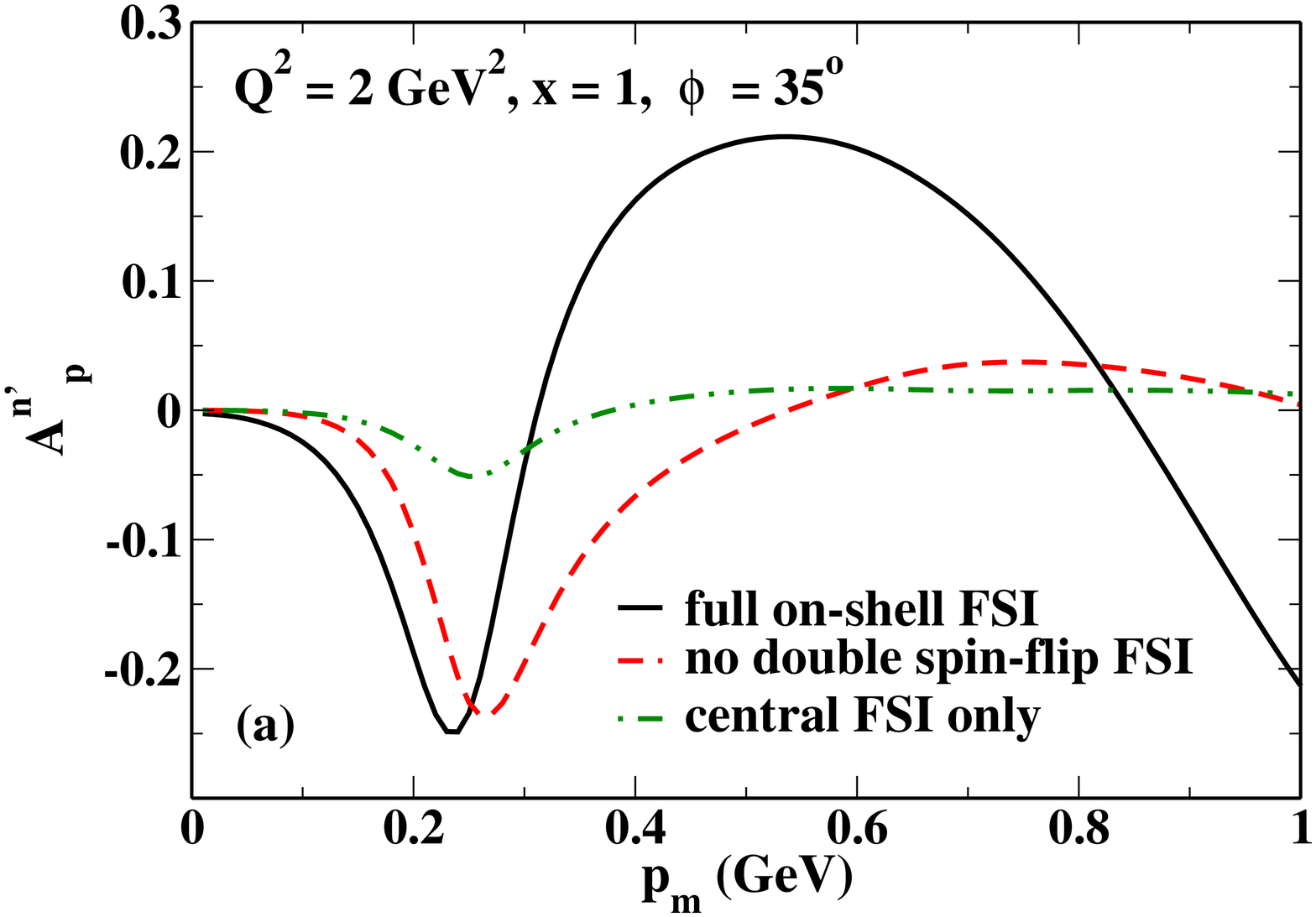}
\includegraphics[width=14pc,angle=270]{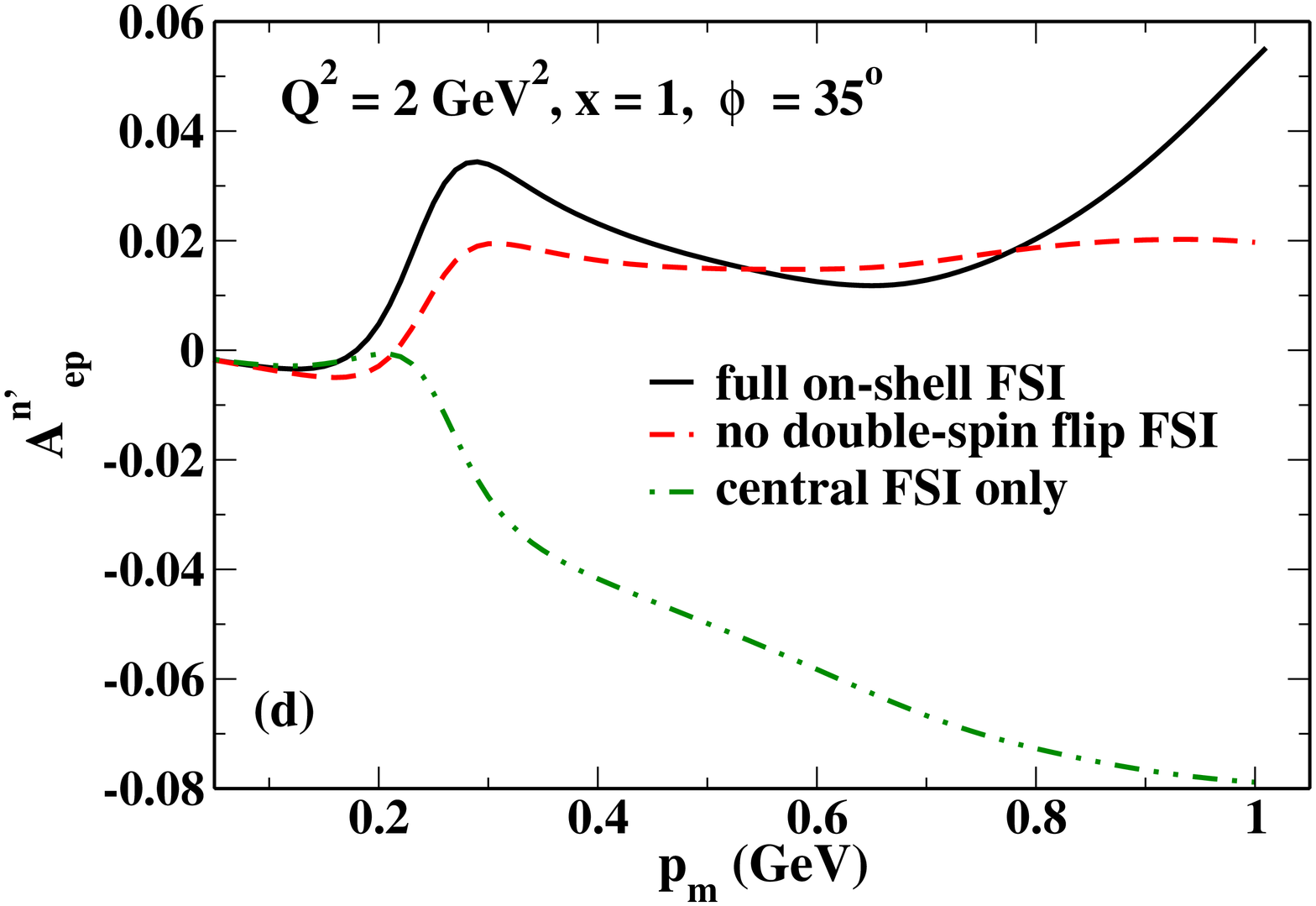}
\includegraphics[width=14pc,angle=270]{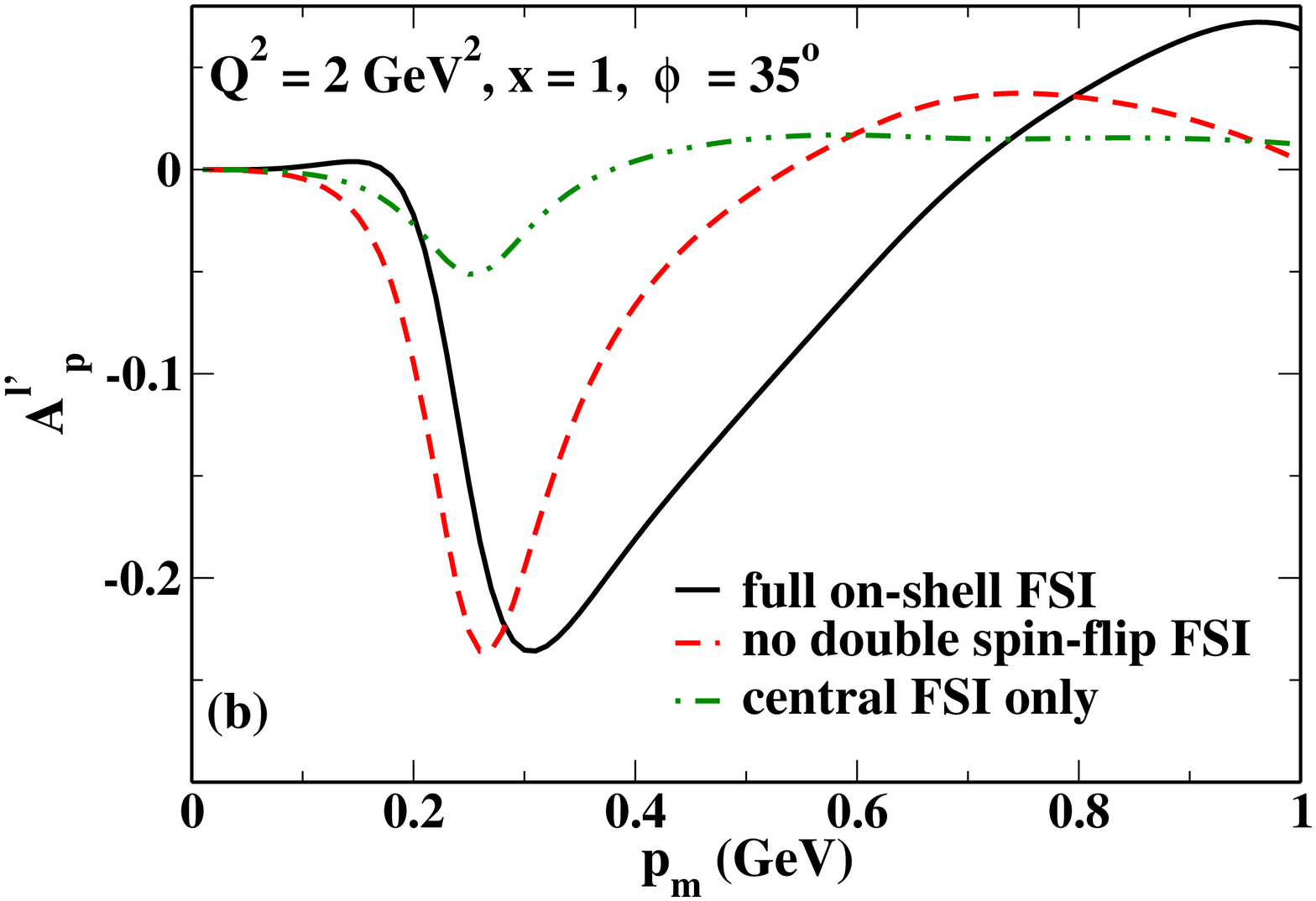}
\includegraphics[width=14pc,angle=270]{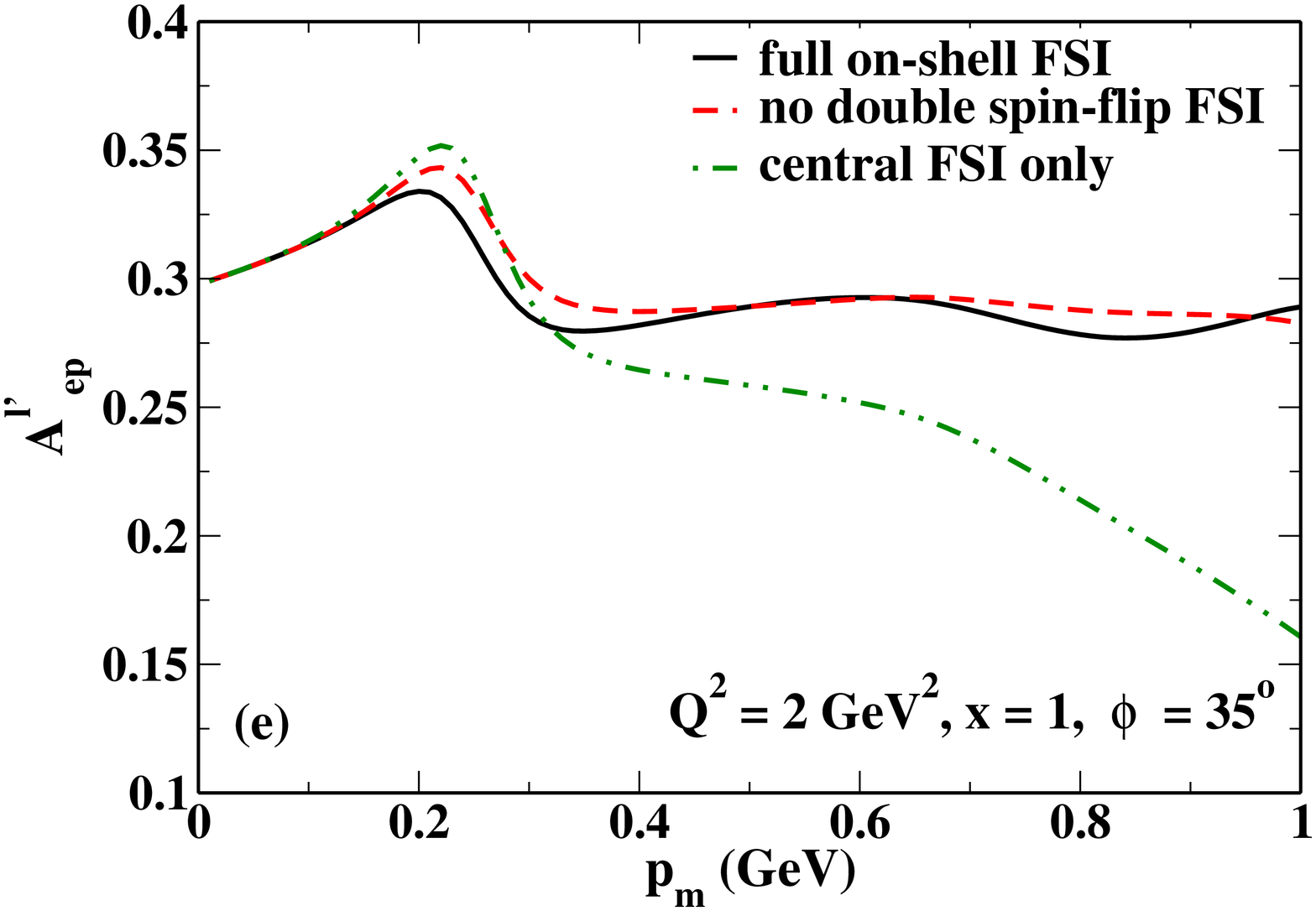}
\includegraphics[width=14pc,angle=270]{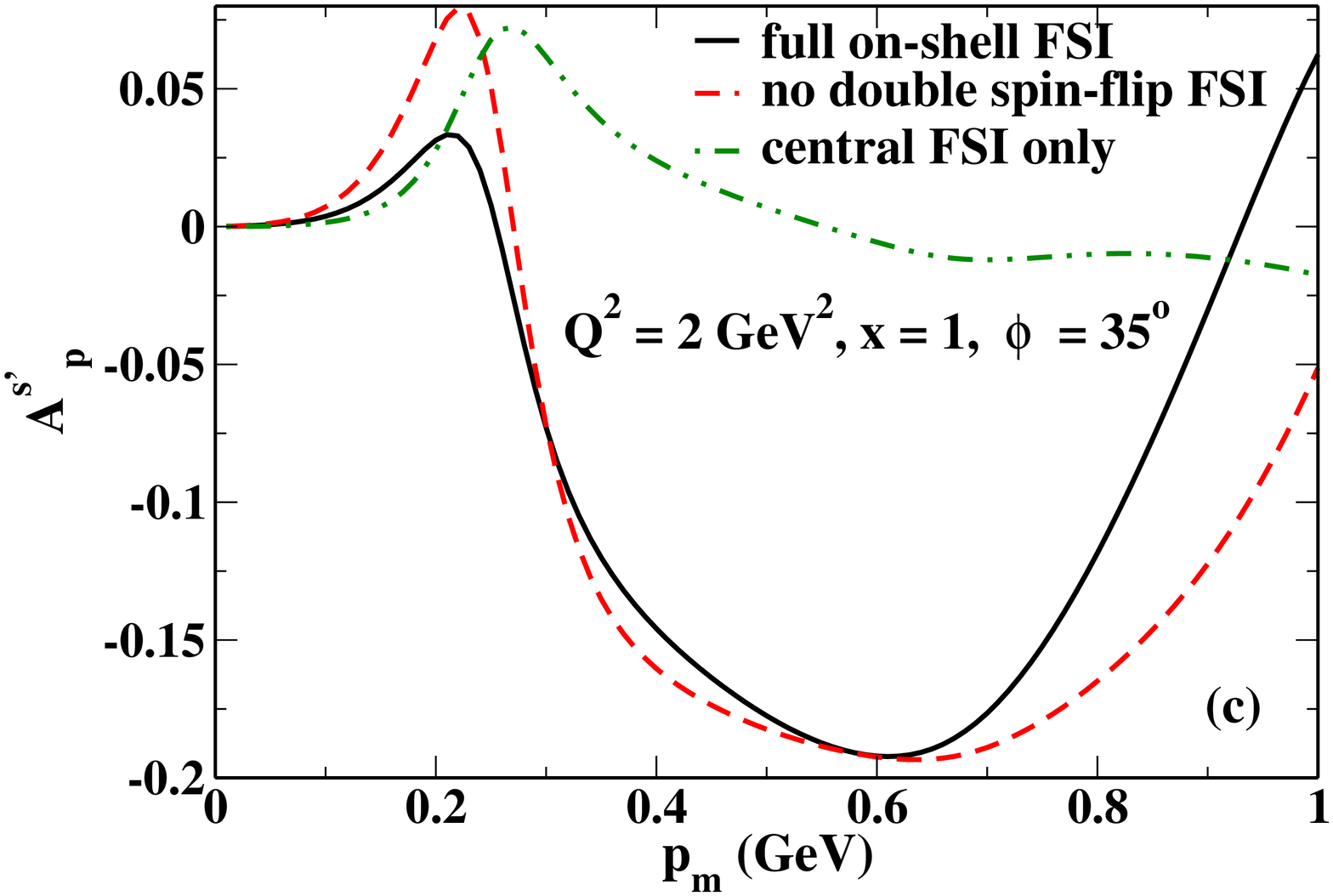}
\includegraphics[width=14pc,angle=270]{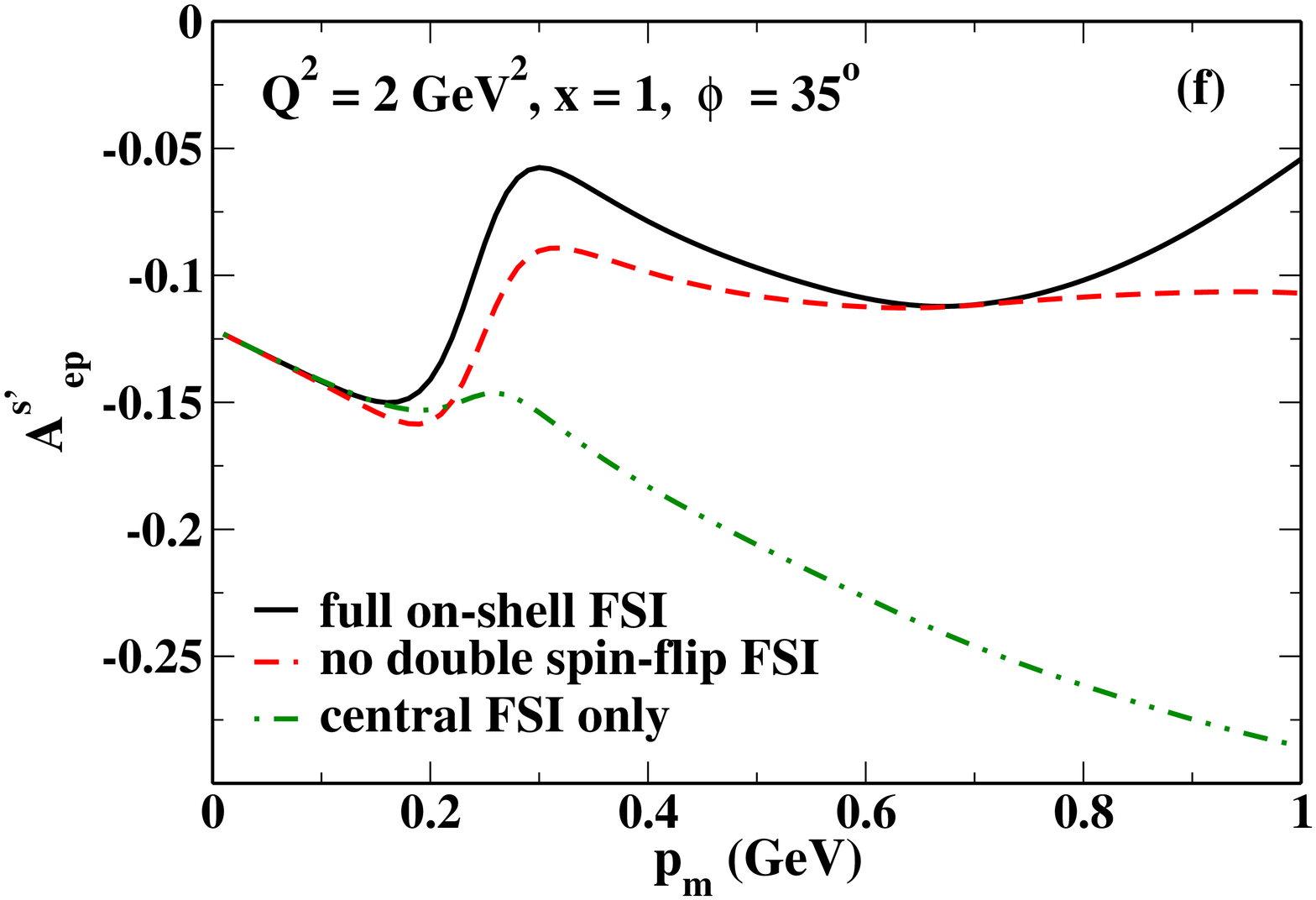}
\caption{(Color online)  The six panels show the six asymmetries plotted versus the missing momentum $p_m$
for a beam energy of $5.5$ GeV, a transferred four-momentum
of $Q^2 = 2$ GeV$^2$, $\phi_p = 35^o$, and $x = 1$. We show $A^{n'}_p$ (panel a)), $A^{l'}_p$ (panel b)), $A^{s'}_p$ (panel c)),
$A^{n'}_{ep}$ (panel d)), $A^{l'}_{ep}$ (panel e)), and $A^{s'}_{ep}$ (panel f)). The curves shown have been calculated
with the full on-shell FSIs (solid), with central and single spin-flip FSIs (dashed), and with central FSI only (dash-dotted). }
\label{fig_spindepfsi}
\end{figure}

In Fig. \ref{fig_spindepfsi}, we show the contributions of the central, central and single spin-flip, and full FSIs to the
six asymmetries at $Q^2 = 2$ GeV$^2$, $x = 1$, and $\phi_p = 35^o$. The clear message from this figure is that
a calculation including only central FSIs will fail completely for missing momenta beyond $p_m = 0.2$ GeV. For the
three asymmetries accessible with an unpolarized beam, the central FSI on its own leads to a mostly zero asymmetry, whereas
spin dependent FSIs lead to large structures in these observables. For the asymmetries accessible with a polarized electron beam
only, the purely central FSI leads to non-zero results for the asymmetries, but the inclusion of spin-dependent FSIs leads to huge
changes, both of the shape and size. The differences between the full FSIs and the FSIs without the three double spin-flip
terms is largest for $A^{n'}_p$, where the double spin-flip contribution leads to a broad bump for missing momenta from
$p_m \approx 0.3$ GeV to $p_m \approx 0.8$ GeV. The double spin-flip contributions to $A^{l'}_p$ are significant as well,
leading to an increase in the magnitude of the asymmetry for medium missing momenta. The double spin effects for
$A^{s'}_p$, are smaller, leading to a reduction in magnitude of a small peak at $p_m \approx 0.2$ GeV and for large missing momenta,
$p_m > 0.8$ GeV. For the asymmetries that require a polarized beam, the influence of the double spin-flip terms is smaller.
There are no huge modifications, and the largest corrections appear for the peak structures around $p_m \approx 0.2$ GeV, and for
very large missing momenta. This figure gives the impression that an additional polarization, i.e. the beam polarization, may
play a similar role as an additional spin-dependence in the FSI, and that at least one of them - either a spin-dependent FSI or
a polarized beam - needs to be present to generate an approximately correct ejectile polarization asymmetry. Once the beam is polarized,
it seems that introducing the additional double spin-flip terms does not really change the results too much.
A very similar picture emerges for $x = 1.3$, away from the quasi-elastic peak. The only difference is that the role of the double spin-flip terms becomes even more important for the asymmetries with an unpolarized beam.

\section{Summary and Outlook}

In this paper, we have introduced a formalism for the calculation of asymmetries relevant to a polarized ejectile
proton in various frames.
In particular, we have presented calculations in a frame relevant to the actual experimental set-up, and avoided any issues with
the definition of projection directions in the case that $\theta_p = 0,\pi$. We have employed a fully relativistic $D(e,e'p)$ calculation in impulse approximation. We have used a parametrization of experimental $NN$ data from SAID to describe the full $pn$ scattering amplitude
for the final state interaction. This leads to certain limits in the kinematics we can access, as these parametrizations
are available only for lab kinetic energies of $1.3$ GeV or less. In our calculations, we have investigated the effects
of the different contributions to the NN scattering amplitude: the central, spin-orbit, and double-spin-flip parts.

The asymmetries accessible with an unpolarized beam are zero in PWIA.
The influence of the FSIs is very large for all six asymmetries. For the three asymmetries where the PWIA results are non-zero,
the FSIs seem to reduce the asymmetries in general, although increases in asymmetries can also be observed for specific kinematics.
We have investigated the role played by off-shell FSIs, and they turn out to be fairly small in most situations, with the exception of
$A^{s'}_p$. This is interesting, as the contributions from off-shell FSIs in unpolarized deuteron scattering  and for a polarized target
were occasionally quite significant.
In practice, it is good news that the off-shell FSIs are small, as any remaining theoretical uncertainty is connected to these terms. So,
the comparison of our theory to data will be very clean.

We have investigated the role played by different parts of the spin-dependent proton - neutron scattering amplitude in the
final state interactions. Using only a central FSI is completely inadequate for all asymmetries. Spin-dependent terms need to be included,
and even the double spin-flip contributions are large, in particular for the asymmetries accessible with an unpolarized beam.

Currently, there are no $D(e,e' \vec p)n$ data at high $Q^2$ available, but an experiment could easily be performed at Jefferson Lab.
In view of the high sensitivity to double spin-flip terms, we feel that one of the most interesting measurements would be
to take data for $A^{n'}_p$ for $x = 1.3$, or even for $x = 1$. Besides this, measuring ejectile polarization asymmetries for kinematics
where unpolarized data or target polarization data are available would allow one to perform a systematic investigation of the reaction
mechanism.

Our description of FSIs is complete, but we are still missing the contributions from $\Delta$ isobars and other meson exchange currents.
Work on these is in progress.

{\bf Acknowledgments}:  We thank Douglas Higinbotham for discussions on experimental
aspects, and for providing us with experimental references. This work was
supported in part by funds provided by the U.S. Department of Energy
(DOE) under cooperative research agreement under No.
DE-AC05-84ER40150 and by the National Science Foundation under
grant No. PHY-0653312.


\begin{thebibliography}{299}

\bibitem{wallyreview}
M.~Garcon and J.~W.~Van Orden,
  Adv.\ Nucl.\ Phys.\  {\bf 26}, 293 (2001).



\bibitem{ronfranz}
  R.~A.~Gilman and F.~Gross,
  J.\ Phys.\ G {\bf 28}, R37 (2002).


\bibitem{sickreview}
  I.~Sick,
  Prog.\ Part.\ Nucl.\ Phys.\  {\bf 47}, 245 (2001).

\bibitem{cteep}
J.~Ryckebusch, W.~Cosyn, B.~Van Overmeire and C.~Martinez,
  Eur.\ Phys.\ J.\  A {\bf 31}, 585 (2007);
J.~Ryckebusch, P.~Lava, M.~C.~Martinez, J.~M.~Udias and
J.~A.~Caballero,
  Nucl.\ Phys.\  A {\bf 755}, 511 (2005);
  P.~Lava, M.~C.~Martinez, J.~Ryckebusch, J.~A.~Caballero and J.~M.~Udias,
  Phys.\ Lett.\  B {\bf 595}, 177 (2004);
L.~L.~Frankfurt, W.~R.~Greenberg, G.~A.~Miller, M.~M.~Sargsian and M.~I.~Strikman,
  Z.\ Phys.\  A {\bf 352}, 97 (1995).


\bibitem{hallbgmn}
J.~Lachniet {\it et al.}  [CLAS Collaboration],
  Phys.\ Rev.\ Lett.\  {\bf 102}, 192001 (2009).

\bibitem{misakneutstar}
  L.~Frankfurt, M.~Sargsian and M.~Strikman,
  Int.\ J.\ Mod.\ Phys.\  A {\bf 23}, 2991 (2008).



\bibitem{bigdpaper}
S.~Jeschonnek and J.~W.~Van Orden,
Phys.\ Rev.\  C {\bf 78}, 014007 (2008).



\bibitem{wallyfranzwf}
F. Gross, J.~W.~Van~Orden and K.~Holinde,
Phys.\ Rev.\ C {\bf 41}, R1909 (1990);
 F. Gross, J. W. Van Orden and K. Holinde,
 Phys.\ Rev. \ C {\bf 45}, 2094
(1992).



\bibitem{said}
R.~A.~Arndt, W.~J.~Briscoe, I.~I.~Strakovsky and R.~L.~Workman,
  Phys.\ Rev.\  C {\bf 76}, 025209 (2007); data available through
SAID, http://gwdac.phys.gwu.edu/




\bibitem{misak}
M.~M.~Sargsian,
  Int.\ J.\ Mod.\ Phys.\  E {\bf 10}, 405 (2001);
M.~M.~Sargsian, T.~V.~Abrahamyan, M.~I.~Strikman and
L.~L.~Frankfurt,
  Phys.\ Rev.\  C {\bf 71}, 044614 (2005);
  L.~L.~Frankfurt, M.~M.~Sargsian and M.~I.~Strikman,
  Phys.\ Rev.\  C {\bf 56}, 1124 (1997).

\bibitem{misakbrandnew}
  M.~M.~Sargsian,
  arXiv:0910.2016 [nucl-th].


\bibitem{genteikonal}
  J.~Ryckebusch, D.~Debruyne, P.~Lava, S.~Janssen, B.~Van Overmeire and T.~Van Cauteren,
  Nucl.\ Phys.\  A {\bf 728}, 226 (2003);
D.~Debruyne, J.~Ryckebusch, W.~Van Nespen and S.~Janssen,
  Phys.\ Rev.\  C {\bf 62}, 024611 (2000);
  B.~Van Overmeire and J.~Ryckebusch,
  Phys.\ Lett.\  B {\bf 650}, 337 (2007);
W.~Cosyn and J.~Ryckebusch,
  arXiv:0904.0914 [nucl-th].



\bibitem{ciofi}
C.~Ciofi~degli Atti and L.~P.~Kaptari,
  Phys.\ Rev.\  C {\bf 71}, 024005 (2005);
C.~Ciofi delgi Atti and L.~P.~Kaptari,
  arXiv:0705.3951 [nucl-th];
C.~Ciofi degli Atti, L.~P.~Kaptari and D.~Treleani,
  Phys.\ Rev.\  C {\bf 63}, 044601 (2001).


\bibitem{laget}
  J.~M.~Laget,
  Phys.\ Lett.\  B {\bf 609}, 49 (2005).


\bibitem{schiavilla}
  R.~Schiavilla, O.~Benhar, A.~Kievsky, L.~E.~Marcucci and M.~Viviani,
  Phys. Rev.  C{\bf 72}, 064003 (2005).


\bibitem{halladata}
Jefferson Lab Experiment E01 - 020, spokespersons W. Boeglin, M. Jones, A. Klein, P. Ulmer,
J. Mitchell, E. Voutier.

\bibitem{egiyansrc}
K.~S.~Egiyan {\it et al.}  [CLAS Collaboration],
  Phys.\ Rev.\ Lett.\  {\bf 96}, 082501 (2006);
K.~S.~Egiyan {\it et al.}  [CLAS Collaboration],
  Phys.\ Rev.\  C {\bf 68}, 014313 (2003).


\bibitem{blast}
BLAST data from MIT Bates, Ph. D. thesis A. Maschinot (MIT 2005).


\bibitem{jerrygexp}
G.~Gilfoyle, spokesperson, Jefferson Lab Hall B, E5 run period;
G.P. Gilfoyle, (the CLAS Collaboration),
'Out-of-Plane Measurements of the Fifth Structure Function of the Deuteron',
Bull. Am. Phys. Soc., Fall DNP Meeting, DF.00010(2006).


\bibitem{wernernewprop}
W. Boeglin, spokesperson, proposal to Jefferson Lab PAC 33, 2007.


\bibitem{targetpol}
  S.~Jeschonnek and J.~W.~Van Orden,
   Phys.\ Rev.\  C {\bf 80}, 054001 (2009).

\bibitem{hudata}
  B.~Hu {\it et al.},
  Phys.\ Rev.\  C {\bf 73}, 064004 (2006).

\bibitem{mainzrecoil}
  D.~Eyl {\it et al.},
  Z.\ Phys.\  A {\bf 352}, 211 (1995).

\bibitem{batesdeuteronrecoil}
  B.~D.~Milbrath {\it et al.}  [Bates FPP collaboration],
  Phys.\ Rev.\ Lett.\  {\bf 80}, 452 (1998)
  [Erratum-ibid.\  {\bf 82}, 2221 (1999)].


\bibitem{protonelasticrecoil_lowqsq}
  G.~Ron {\it et al.},
  Phys.\ Rev.\ Lett.\  {\bf 99}, 202002 (2007).

\bibitem{protonelasticrecoil_highqsq}
  M.~K.~Jones {\it et al.}  [Jefferson Lab Hall A Collaboration],
  Phys.\ Rev.\ Lett.\  {\bf 84}, 1398 (2000);
  O.~Gayou {\it et al.}  [Jefferson Lab Hall A Collaboration],
  Phys.\ Rev.\ Lett.\  {\bf 88}, 092301 (2002).




  \bibitem{fpp_technical}
  J.~Glister {\it et al.},
  Nucl.\ Instrum.\ Meth.\  A {\bf 606}, 578 (2009).


\bibitem{batescarbonrecoil}
  R.~J.~Woo {\it et al.},
  Phys.\ Rev.\ Lett.\  {\bf 80}, 456 (1998).

\bibitem{photodisrecoil}
  X.~Jiang {\it et al.}  [Jefferson Lab Hall A Collaboration],
  Phys.\ Rev.\ Lett.\  {\bf 98}, 182302 (2007).


\bibitem{kellypionprod}
  J.~J.~Kelly {\it et al.},
  Phys.\ Rev.\  C {\bf 75}, 025201 (2007).



\bibitem{raskintwd}
A. S. Raskin and T. W. Donnelly, {\sl Ann. of Phys.} {\bf 191}, 78
(1989).


\bibitem{dmtrgross}
V. Dmitrasinovic and F. Gross, {\sl Phys. Rev. C} {\bf40}, 2479
(1989).



\bibitem{grosseqn}
F. Gross, Phys. Rev. {\bf 186}, 1448 (1969); Phys. Rev. D  {\bf 10},
               223 (1974), Phys. Rev. C {\bf 26}, 2203 (1982).








\end{thebibliography}
\end{document}